\documentclass[prc,preprint,showpacs,tightenlines,%
                floatfix,amsfonts,nofootinbib]{revtex4}
\usepackage{graphicx}  
\usepackage{bm}  
\usepackage{amssymb}

\begin{document}

%
% Begin: Simple substitution macros used in the text
%
%%%  syntax:
%
%\newcommand{macro name}[arguments]{what it means}
%

\newcommand{\finalnewpage}{\newpage}
\newcommand{\newg}{{\skew2\overline g}_\rho}
\newcommand{\Nbar}{\skew3\overline N \mkern2mu}
\newcommand{\stroke}[1]{\mbox{{$#1$}{$\!\!\!\slash \,$}}}

%
% End: Simple substitution macros
%

% Other definitions:

\newcommand{\beq}{\begin{equation}}
\newcommand{\eeq}{\end{equation}}
\newcommand{\beqa}{\begin{eqnarray}}
\newcommand{\eeqa}{\end{eqnarray}}

% Macro for taking limits underneath arrows:
% use this one for automatic arrow length,
% or write it explicitly in the formula
\def\Inthelimit#1{\lower1.9ex\vbox{\hbox{$\  
   \buildrel{\hbox{\Large \rightarrowfill}}\over{\scriptstyle{#1}}\ $}}}
%
%

%\title{The Axial-Vector Current in Nuclear Many-Body Physics}
%\title{The Axial$\,$-Vector Current in Nuclear Many-Body Physics}
%\title{The Axial-$\!$Vector Current in Nuclear Many-Body Physics}
\title{The Axial$\mkern1.2mu$-$\mkern-2.1mu$Vector Current %
        in Nuclear Many-Body Physics}

\author{Sergei M. Ananyan}\email{s.ananyan@megaputer.com}
\author{Brian D. Serot}\email{serot@iucf.indiana.edu}
\affiliation{Department of Physics and Nuclear Theory Center
             Indiana University, Bloomington, IN\ \ 47405}
\author{John Dirk Walecka}\email{walecka@physics.wm.edu}
\affiliation{Department of Physics, 
             The College of William and Mary,\
             Williamsburg, VA\ \ 23187}

%%%%  The following lines FOOL REVTeX_4 into leaving some space
%     before the date in preprint mode.
%     
\author{\null}
\noaffiliation

% use this for the draft version
%\date{\today ;\ \ {\bf DRAFT}}
%
% use one of these for the final version
%\date{\today}    
\date{September, 2002}

\begin{abstract}
Weak-interaction currents are studied in a recently proposed
effective field theory of the nuclear many-body problem.
The Lorentz-invariant effective field theory contains nucleons,
pions, isoscalar scalar ($\sigma$) and vector ($\omega$) fields, and
isovector vector ($\rho$) fields.
The theory exhibits a nonlinear realization of $SU(2)_L \times
SU(2)_R$ chiral symmetry and has three desirable features:
it uses the same degrees of freedom to describe the axial-vector current
and the strong-interaction dynamics,
it satisfies the symmetries of the underlying theory of quantum
chromodynamics, and
its parameters can be calibrated using strong-interaction phenomena,
like hadron scattering or the empirical properties of finite nuclei.
Moreover, it has recently been verified that for normal nuclear
systems, it is possible to systematically
expand the effective lagrangian in powers of the meson fields (and their 
derivatives) and to reliably truncate the expansion
after the first few orders.
Here it is shown that the expressions for the axial-vector
current, evaluated through the first few orders in the field
expansion, satisfy both PCAC and the Goldberger--Treiman relation,
and it is verified that the corresponding vector and axial-vector
charges satisfy the familiar chiral charge algebra.
Explicit results are derived for the Lorentz-covariant, axial-vector,
two-nucleon amplitudes, from which axial-vector meson-exchange
currents can be deduced.
\end{abstract}

\smallskip
\pacs{24.10.Cn, 11.40.Ha, 24.10.Jv, 12.40.Vv.}

\maketitle

\section{Introduction}
%\label{sec:intro}

Although quantum chromodynamics (QCD) is known to be the fundamental theory   
of the strong interaction, it is much more efficient to use hadronic degrees
of freedom to describe few- and many-body nuclear systems at low energies.
Considerable effort over the last 30 years has shown that models based
on baryons and mesons can provide a realistic description of the 
nucleon--nucleon (NN) interaction, nuclear matter saturation, and the bulk
and single-particle properties of finite nuclei.
(For reviews, see, for example, 
Refs.~\cite{Mac89,Mac00,Wal95,Ser86,Ser97}.)
The hadrons are described by effective fields whose interactions are
determined by a local, Lorentz-invariant lagrangian.
The modern viewpoint \cite{Wei79,Wei95}
is that the resulting relativistic, quantum, effective
field theory provides the most general way to parametrize an $S$ matrix
(or other observables \cite{FHT01}) 
that is consistent with the constraints of quantum
mechanics, special relativity, unitarity, causality, cluster decomposition,
and the desired internal symmetries.
Thus there is no reason that relativistic quantum field theory should be
reserved for ``elementary'' particles only.
We will refer to Lorentz-covariant, relativistic effective
field theories based on hadrons as
{\em quantum hadrodynamics} or QHD \cite{Wal74,Ser86,Ser97,Com00,Ser00}.

It has also been known for many years that an accurate description of 
electroweak interactions in nuclei requires a consideration of mesonic
degrees of freedom (i.e., ``fields'') in addition to the nucleons.
(For a review, see Refs.~\cite{Dub76,Rho79,Ris89}.
An example of more recent work is Ref.~\cite{Dmi98}.)
The mesons are responsible for the forces between the nucleons and also lead
to meson-exchange currents (MEC) that contribute to electroweak processes.
A desirable theory of the axial-vector current
should satisfy the following three conditions:
\begin{itemize}
\item
It should use the same degrees of freedom to describe the axial current
and the strong-interaction dynamics 
($\pi$N scattering, NN scattering, and nuclear structure);
the basic phenomenological features of the latter are well known.
\item
It should satisfy the same internal symmetries as the underlying theory
of QCD: the discrete symmetries of the strong interaction and (approximate)
isospin and chiral symmetries, with the last being spontaneously broken.
The enforcement of the continuous symmetries is necessary to ensure the
conservation of the vector, isovector current (CVC) and the partial 
conservation of the axial-vector, isovector current (PCAC).
\item
It should be possible to calibrate the parameters of the theory using
strong-interaction phenomena,
like $\pi$N scattering and the properties of finite nuclei, so that one
can deduce well-defined and unambiguous currents to be used
in the calculation of electroweak processes.
This is especially important in effective field theories, because
these contain all (non-redundant) interaction terms that are consistent
with the underlying symmetries \cite{Wei95,Ser97}.
\end{itemize}

In discussing MEC for the weak interaction, CVC implies that the vector part
of the current can be determined by performing an isospin rotation on the
isovector part of the electromagnetic current \cite{Wal95}.
Thus the vector parts of the weak 
{\em exchange\/} currents can be derived from
the isovector part of the corresponding electromagnetic 
{\em exchange\/} currents,
which have been determined accurately over the last two decades.
In contrast, the axial-vector parts of the exchange currents require 
additional theoretical input.

There have been many attempts to describe axial-vector exchange currents
(AXC) in models based on hadronic degrees of freedom.
An important, early contribution was made by Kubodera, Delorme,
and Rho \cite{Kub78}, who predicted the dominant long-range
piece of the AXC using
current algebra and the assumption of pion-exchange dominance, which yield
the leading-order terms in an expansion in inverse powers of the nucleon
mass.
Some interesting recent approaches include: a description of AXC using all
the degrees of freedom contained in the phenomenological NN 
potential \cite{Blu92,Kir92,Kir94,Ris94},
a model based on ``hard pions'' \cite{Iva79,Tow92,Con96,Sme97}, 
and the application of chiral perturbation theory 
(ChPT) \cite{Rho91,Par93,Par94,Par96}.

While these approaches enjoyed several successes \cite{Lee93,Ris94}, 
they all contain shortcomings that make them undesirable, 
at least according to the criteria presented above.
In particular, current-algebra techniques are difficult to extend to
multi-pion contributions or to mesons other than the pion.
Models based
on a phenomenological NN potential do not explicitly incorporate chiral
symmetry and do not explicitly include AXC that arise from the direct
interaction of the exchanged mesons with the axial current.
Hard-pion models use a linear realization of the chiral symmetry; thus,
inclusion of the $\rho$ meson necessitates the inclusion of its chiral
partner, the $a_1$, which is known to be relatively unimportant in the NN
interaction and in nuclear structure~\cite{Ser92}.
Moreover, the strong, mid-range NN attraction, which arises predominantly 
from correlated two-pion exchange, and which is crucial for an accurate
description of NN phase shifts and nuclear matter 
properties \cite{Mac89,Ser97,Com00,LARGE}, 
is difficult to generate in models with
linear chiral symmetry without also generating 
unrealistic many-nucleon forces \cite{Fur96}.
In contrast, when the
chiral symmetry is realized {\em nonlinearly}, the mid-range attraction
can be efficiently simulated by introducing an effective, scalar, 
isoscalar, chiral singlet
$\sigma$ meson with a mass of roughly 500 MeV~\cite{TANG95,Fur97}.

Finally, models based on ChPT attempt to explain all of the dynamics using
nucleons and pions alone; this makes a description of the NN interaction
and of nuclear structure very complicated, since one must generate much of
the strong-interaction dynamics using multi-pion loop 
processes~\cite{Epe98,Epe00,Epe02,Ent02,Ent02a}.
These shortcomings motivate the search for alternative descriptions
of the AXC.

In Ref.~\cite{Ana98}, a lagrangian-based model that contains $\pi$,
$\sigma$, and $\omega$ meson fields was used to construct the AXC.
This model incorporates the desirable qualities enumerated earlier: it
contains the mesons responsible for the dominant features of the NN
interaction, it respects both isospin symmetry and (approximate)
spontaneously broken chiral symmetry, and (in principle) its
parameters can be calibrated to the properties of nuclei by using the
mean-field approximation for the meson fields.  Nevertheless, the
implementation of the chiral symmetry in this model (which is based on
the well-known Sigma model \cite{Sch57,GML60,Wal95}) 
turns out to be too restrictive.  
In particular, it is impossible to satisfy both PCAC and the
Goldberger--Treiman relation without destroying the familiar chiral
charge algebra.  It is also impossible to reproduce the empirical
equilibrium point of nuclear matter in the mean-field approximation
\cite{Ker74,Fur96}.  Moreover, the linear realization of the chiral
symmetry makes it cumbersome to include the $\rho$ meson \cite{Ser92}.
A viable description of the AXC requires a more general implementation of
the chiral symmetry.

In the present work, we derive the axial-vector current
using a recently proposed
QHD lagrangian \cite{Fur97,Ser97} that contains nucleons and $\pi$,
$\sigma$, $\omega$, and $\rho$ mesons.  This lagrangian has a {\em
linear\/} realization of the $SU(2)_V$ isospin symmetry and a {\em
nonlinear\/} realization of the spontaneously broken $SU(2)_L \times
SU(2)_R$ chiral symmetry (when the pion mass is zero).  It was shown
in Refs.~\cite{Fur97,PARAMS,Ser00} that by using 
Georgi's naive dimensional analysis
(NDA) \cite{Geo93} and the assumption of {\em naturalness} (namely,
that all appropriately defined, dimensionless couplings are of order
unity), it is possible to truncate the lagrangian at terms
involving only a few powers of the meson fields and their derivatives,
at least for systems at normal nuclear densities.  It was also shown
that a mean-field approximation to the lagrangian could be interpreted
in terms of density functional theory \cite{Dre90,Ser97,Ser02}, so that
calibrating the parameters to observed bulk and single-particle
nuclear properties incorporates (approximately) many-body effects that
go beyond mean-field theory.\footnote{%
Since our lagrangian contains the
four most important mesons used in boson-exchange models of the NN
interaction \protect\cite{Mac89,Mac00}, 
as well as all of the relevant
meson--nucleon couplings, it should also provide a reasonable
description of the NN data, albeit with different parameter values
\cite{Gro90}.  Thus the parameters could also be sensibly calibrated to NN
data, which should be more useful for studying exchange currents in
few-nucleon systems.}  Explicit calculations of closed-shell nuclei
provided such a calibration and verified the naturalness 
assumption \cite{PARAMS}.
{\em This approach therefore embodies the 
three desirable features needed for a description of the axial-vector 
current in the nuclear many-body problem.}

This effective field theory (EFT) also overcomes the difficulties 
found in Ref.~\cite{Ana98}. 
The chiral symmetry guarantees that the PCAC condition will hold
for the one- and two-body axial currents and pion-production amplitudes,
as we demonstrate explicitly below.
Moreover, the $\pi$N coupling strength $g_A$ enters as a 
{\em free parameter\/} that
can be chosen so that the Goldberger--Treiman relation is satisfied at the
tree level, without any rescaling of the fields.
This result, together with the chiral symmetry, ensures that the conserved
vector and axial-vector charges obey the familiar algebra.
Thus all three required constraints can be satisfied simultaneously in the
present framework.
Moreover, the pion--pion and pion--nucleon parts of the lagrangian 
are exactly the same as those of chiral perturbation 
theory~\cite{Gas88,Ell98,Bira}.

It is important to note that in our EFT, only the pions and nucleons
(the stable particles) can appear on external lines with 
{\em timelike\/} four-momenta.
The heavy non-Goldstone bosons appear only on internal lines (with
{\em spacelike\/} four-momenta) 
and allow us to parametrize the medium- and
short-range parts of the nucleon--nucleon interaction,
as well as the electromagnetic form 
factors of the hadrons~\cite{TANG95,Fur97}.
The heavy bosons are also convenient degrees of freedom for 
describing nonvanishing expectation values of bilinear nucleon
operators, like $\Nbar N$ and $\Nbar \gamma^{\mu} N$, which are
important in nuclear many-body systems~\cite{Ser86,Ser97}.

The remainder of this paper is organized as follows.  In
Sec.~\ref{sec:EFT}, the effective QHD lagrangian is introduced and
described briefly, and the vector and axial-vector currents are
derived for the interaction terms that contain only one derivative of
the pion field.  The currents arise directly from Noether's theorem
and contain the pion field to all orders.  In Sec.~\ref{sec:amps}, the
matrix elements of the one- and two-body axial currents and the
pion-production amplitude arising from these contributions are
computed, and PCAC and the Goldberger--Treiman relation are verified.
It is also demonstrated that in the chiral limit, the conserved
charges corresponding to these currents reproduce the desired algebra.
Sections \ref{sec:nuthree} and \ref{sec:moreamps} extend the analysis
to terms in the lagrangian that are bilinear in derivatives of the
pion field, and Sec.~\ref{sec:rho}
discusses the leading contributions containing a
$\rho$ meson.  The isoscalar $\sigma$ and $\omega$
meson contributions are
given in Sec.~\ref{sec:sigma-omega}. 
Sec.~\ref{sec:summary} contains a summary.

\section{\label{sec:EFT}Effective Field Theory lagrangian}

The effective field theory (EFT) lagrangian considered in the present
paper was proposed in Ref.~\cite{Fur97}.
As discussed in that paper, the nonlinear chiral lagrangian can be
organized in increasing powers of the fields and their derivatives.
To each interaction term we assign an index
\begin{equation}
 \nu \equiv d + {n \over 2} + b \ ,
\end{equation}
where $d$ is the number of derivatives, $n$ is the number of nucleon fields,
and $b$ is the number of non-Goldstone boson fields in the interaction term.
Derivatives on the nucleon fields are not counted in $d$ because they will
typically introduce powers of the nucleon mass $M$, which will not lead to
small expansion parameters~\cite{Fur97}.

It was shown in Refs.~\cite{Fur96,Fur97}
that for finite-density applications
at and below nuclear matter equilibrium density, one can truncate the
effective lagrangian\footnote{%
Two terms of order $\nu = 5$ involving bilinear derivatives of the $\sigma$
and $\omega$ fields are of minor, but nonnegligible, numerical importance
and were included in Ref.~\protect\cite{Fur97}. 
We will not be concerned with such details in this work.}
at terms with $\nu \leq 4$.
It was also argued that by making suitable definitions of the nucleon and
meson fields, it is possible to write the lagrangian in a ``canonical'' form
containing familiar noninteracting terms for all fields, Yukawa couplings
between the nucleon and meson fields, and nonlinear meson 
interactions~\cite{FHT01}.
See Refs.~\cite{Ser97,Fur97} for a more complete discussion.

If we keep terms with $\nu \leq 4$, the chirally invariant lagrangian 
can be written as\footnote{%
We use the conventions of Refs.~\protect\cite{Ser86,Fur97,Ser97}.}
\begin{eqnarray}
{\cal L}_{\mathrm{EFT}} & = & {\cal L}_N  + {\cal L}_4
        + {\cal L}_M  \nonumber \\[7pt]
 & = & \Nbar \left( i {\gamma}^{\mu} \left[ {\partial}_{\mu} + i v_{\mu}
+ i g_{\rho} {\rho}_{\mu} + i g_v V_{\mu} \right]
+ {g_A}\mkern2mu
  {\gamma}^{\mu} {\gamma}_{5} a_{\mu} - M + g_s \phi \right) N 
        \nonumber \\[4pt]
 & & \quad
{} -  { {f_{\rho} g_{\rho}} \over {4 M} } 
\Nbar {\rho}_{\mu \nu} {\sigma}^{\mu \nu} N 
- { {f_{v} g_{v}} \over {4 M} } 
\Nbar {V}_{\mu \nu} {\sigma}^{\mu \nu} N 
- { { {\kappa}_{\pi} } \over M } 
\Nbar {v}_{\mu \nu} {\sigma}^{\mu \nu} N 
+ { {4 {\beta}_{\pi}} \over M } \Nbar N \, {\rm Tr} 
\left( a_{\mu} a^{\mu} \right) \nonumber \\[4pt]
 & & \quad
 {} + {\cal L}_4
+  { 1 \over 2 } \, 
{\partial}_{\mu} \phi \, {\partial}^{\mu} \phi 
+  { 1 \over 4 } f^2_{\pi} \,
{\rm Tr} \left({\partial}_{\mu} U {\partial}^{\mu} U^{\dagger} \right)
- { 1 \over 2 } \, {\rm Tr} 
      \left( {\rho}_{\mu \nu} {\rho}^{\mu \nu} \right) 
- { 1 \over 4 }\, 
{V}_{\mu \nu} {V}^{\mu \nu} 
\nonumber \\[4pt]
 & &\quad
 {} - g_{\rho \pi \pi} { {2 f^2_{\pi}} \over { m^2_{\rho} } } \,
{\rm Tr} \left( {\rho}_{\mu \nu} {v}^{\mu \nu} \right)
+ { 1 \over 2 } \left( 1 + {\eta}_1 { {g_s \phi} \over M } 
+ {{\eta}_2 \over 2} { {g^2_s {\phi}^2} \over {M^2} } \right)
m^2_v V_{\mu} V^{\mu} 
 + { 1 \over {4!} } \,{\zeta}_0 g_v^2 {\left( V_{\mu} V^{\mu} \right)}^2
\nonumber \\[4pt]
 & &\quad
+ \left( 1 + {\eta}_{\rho} \, { {g_s \phi} \over M } \right)
m^2_{\rho} \, {\rm Tr} \left( {\rho}_{\mu} {\rho}^{\mu} \right)
-  m^2_s {\phi}^2 \left( { 1 \over 2 } + { {{\kappa}_3 } \over {3!} }
{ {g_s \phi} \over M } + { {{\kappa}_4 } \over {4!} } 
{ {g^2_s {\phi}^2} \over {M^2} }\right) ,
\label{eq:eft-lagrangian}
\end{eqnarray}
where the nucleon, pion, sigma, omega, and rho fields are denoted by
$N$, $\mbox{\boldmath $ \pi $}$, $\phi$, $V_\mu$, and $\rho_\mu \equiv
{ 1 \over 2 } \, \mbox{\boldmath $ \tau  \! \cdot \! \rho $}_{\mu}$, 
respectively, 
$V_{\mu\nu} \equiv \partial_\mu V_\nu - \partial_\nu V_\mu$, and
${\sigma}^{\mu \nu} \equiv {i \over 2} [{\gamma}^{\mu}, {\gamma}^{\nu}]$.
The trace ``Tr'' is in the $2 \times 2$ isospin space.
The pion field enters through the combinations
\begin{eqnarray}
U & \equiv & \exp(i \mbox{\boldmath $ \tau  \! \cdot \! \pi $} / f_\pi ) \ , 
\qquad \qquad
 \xi  \equiv  \exp(i \mbox{\boldmath $ \tau  \! \cdot \!
\pi $ } / 2 f_\pi ) \ , \label{eq:Upi} \\[5pt]
a_{\mu} &\equiv & - {i \over 2} \left( {\xi}^{\dag} {\partial}_{\mu} {\xi} -
\xi {\partial}_{\mu} {\xi}^{\dag} \right) \label{eq:a-mu} \\
&\approx& { 1 \over {2 f_{\pi}} } \, \mbox{\boldmath $ {\tau} 
\cdot $}\,{\partial}_{\mu} \mbox{\boldmath $ \pi $} + { 1 \over {12
f_{\pi}^3} } \left[ \mbox{\boldmath $ \pi \cdot $} \left( {\partial}_{\mu}
\mbox{\boldmath $ \pi $} \right)
  \, \mbox{\boldmath $ \tau \! \cdot \! \pi$} 
- {\pi}^2 \, \mbox{\boldmath $ {\tau}
 \cdot $}\,{\partial}_{\mu} \mbox{\boldmath $ \pi $} 
\right] + O({\pi}^4 \partial_\mu \pi) \ , \\[5pt]
v_{\mu} &\equiv&  - {i \over 2} \left( {\xi}^{\dag} {\partial}_{\mu} {\xi} +
\xi {\partial}_{\mu} {\xi}^{\dag} \right) \label{eq:v-mu} \\
&\approx & { 1 \over {4 f_{\pi}^2} } \, \mbox{\boldmath $ \tau  \,
\cdot $} \left[ \mbox{\boldmath $ \pi $} 
   \bm{\times} {\partial}_{\mu}
\mbox{\boldmath $ \pi $} \right] + O({\pi}^3 \partial_\mu \pi) 
  \ , \\[7pt]
v_{\mu \nu} &\equiv & \partial_\mu v_\nu - \partial_\nu v_\mu + i[v_\mu , 
v_\nu ] = - i [a_{\mu},a_{\nu}] \label{eq:v-mu-nu} \\[5pt]
&\approx & { 1 \over {2 f_{\pi}^2} } \, \mbox{\boldmath $ \tau \,
\cdot $} \left[ \partial_\mu
\mbox{\boldmath $ \pi $} \bm{\times} {\partial}_{\nu}
\mbox{\boldmath $ \pi $} \right] + 
 O[{\pi}^2 (\partial_\mu \pi ) (\partial_\nu \pi )] \ .
\end{eqnarray}
The rho meson enters through the covariant field tensor
\begin{equation}
{\rho}_{\mu \nu}=D_{\mu} {\rho}_{\nu} - D_{\nu} {\rho}_{\mu} +i \, \newg
[{\rho}_{\mu}, {\rho}_{\nu}] \ ,
\label{eq:rhofieldtensor}
\end{equation}
where the covariant derivative is defined by
\begin{equation}
D_{\mu} {\rho}_{\nu} \equiv {\partial}_{\mu}{\rho}_{\nu} + i [v_{\mu},
{\rho}_{\nu}] \ ,
\label{eq:rhoderiv}
\end{equation}
and $\newg$ is a free parameter~\cite{Ser97,Fur97}.
${\cal L}_4$ contains $\pi\pi$ and $\pi$N interactions of order
$\nu = 4$ that are not needed in this work.
A numerically insignificant $\nu = 4$ term proportional to
$\phi^2 \,{\rm Tr} \left( {\rho}_{\mu} {\rho}^{\mu} \right)$ has been 
omitted.

This EFT lagrangian provides a consistent framework for 
explicitly calculating the two-body exchange currents originating from
meson--nucleon interactions in nuclei. 
According to naive dimensional analysis (NDA), all of the coupling 
parameters are written in dimensionless form and should be of order unity,
if the theory obeys naturalness; this has been verified for the parameters
that are relevant for mean-field nuclear structure 
calculations~\cite{Fur97,PARAMS}.
Moreover, all the constants entering the
lagrangian~(\ref{eq:eft-lagrangian}) are assumed to be determined from
calibrations to nuclear and nucleon structure data, hadronic decays,
and $\pi$N scattering observables~\cite{Fur97,Ell98}.

The familiar axial-vector, exchange-current results are reproduced 
by the $\nu = 2$ terms, when they are expanded to leading
order in the pion field. 
In addition, the $\nu =3$ terms lead to new contributions to the
axial-vector current that will be calculated explicitly and
shown to preserve the correct charge algebra. 

The same sequence of steps is used to calculate
the axial-vector current from both the 
$\nu = 2$ and $\nu =3$ terms. 
First, we calculate canonical
momenta and identify the Noether vector and axial-vector currents for
the effective lagrangian to a given order in $\nu$. 
Next, we demonstrate
that the correct chiral charge algebra holds {\em to
all orders in the pion field\/} in each case (except for contributions
involving the $\rho$ meson, simply in the interest of brevity).
This result differs from that obtained
in the $\sigma\omega$ model with a linear representation of the chiral
symmetry. 
Here it follows as a direct consequence of the {\em nonlinear}
implementation of the chiral symmetry in the EFT lagrangian. 
One then calculates for all contributions the one-body matrix elements of
the axial current, the axial-current, pion-production amplitude on 
a single nucleon, and 
the two-body matrix elements of the axial current.
PCAC is verified for each of these
amplitudes, the soft-pion limit is investigated for pion production,
and the two-body, axial-current matrix elements can be used to identify 
the corresponding nuclear AXC.

\subsection{The pionic part of the lagrangian}

We begin by considering the $\pi$N part of the EFT lagrangian.
This consists of a purely pionic part
and a part that involves pions interacting with nucleons. 
We will first consider the $\nu = 2$ terms 
in the NDA counting scheme and later analyze contributions from
additional terms with $\nu = 3$.  
The $\nu = 2$ terms involving only pions and nucleons can be written 
as\footnote{%
The $\phi$ and $V^{\mu}$ fields are isoscalar chiral singlets, and the
terms in ${\cal L}_2$ involving these fields do not contribute to the
currents.}
\begin{equation}
{\cal L}_{2}= {\cal L}_{\pi} + {\cal L}_{\pi N} \ ,
\label{eq:lagrangian-nu2}
\end{equation}
where
\begin{equation}
{\cal L}_{\pi}= { {f_{\pi}^2} \over 4} \left\{ {\rm Tr}
\left({\partial}_{\mu} U {\partial}^{\mu} U^{\dag} \right)+ m_{\pi}^2
\, {\rm Tr} \left( U + U^{\dag} - 2 \right) \right\}
\label{eq:lagrangian-pi}
\end{equation}
and
\begin{equation}
{\cal L}_{\pi N}=\Nbar \left\{ i {\gamma}^{\mu} \left
[ {\partial}_{\mu} + {1 \over 2} \left({\xi}^{\dag} {\partial}_{\mu}
{\xi} + {\xi} {\partial}_{\mu} {\xi}^{\dag} \right) \right] - {i \over
2} \, {g_A}\mkern2mu
{\gamma}^{\mu} {\gamma}_5 \left( {\xi}^{\dag}
{\partial}_{\mu} {\xi} - {\xi} {\partial}_{\mu} {\xi}^{\dag} \right) -
M \right\} N \ ,
\label{eq:lagrangian-pi-N}
\end{equation}
with $U$ and $\xi$ defined in Eq.~(\ref{eq:Upi}).

In this lagrangian, 
the $SU(2)_V$ isovector symmetry is represented linearly,
while the $SU(2)_L \times SU(2)_R$ chiral
symmetry is realized in a nonlinear fashion~\cite{Col69,Cal69}.
Transformations of the fields are defined by
\begin{eqnarray}
U(x) & \rightarrow & LU(x)R^{\dag} \ , \nonumber \\[6pt]
{\xi}(x) & \rightarrow & L
{\xi}(x) h^{\dag}(x)= h(x) {\xi}(x) R^{\dag} \ , 
   \nonumber \\[6pt]
N(x) & \rightarrow & h(x)N(x) \ .
\label{eq:chiral-trans}
\end{eqnarray}
A vector $SU(2)_V$ transformation with infinitesimal group parameters
$\mbox{\boldmath $ \beta $}$ is specified by
\begin{equation}
L  =  {\rm exp}(i \mbox{\boldmath $ \beta  \cdot \tau$} /2 )= R = h(x) \ ,
\end{equation}
while an axial transformation of these fields with
infinitesimal parameters $\mbox{\boldmath $ \alpha $}$ is given by
\begin{equation}
  L  =  {\exp} (i \mbox{\boldmath $ \alpha \cdot \tau$}/2) \ , \qquad
  R  =  {\exp} ({-i} \mbox{\boldmath $ \alpha \cdot \tau$}/2) \ , \qquad
  h(x) = {\exp}[i \mbox{\boldmath $ \gamma $}(x)
      \mbox{\boldmath${}\cdot \tau $} / 2] \ .
\end{equation}
The isovector function $\mbox{\boldmath $\gamma$}(x) $ is defined 
implicitly by Eq.~(\ref{eq:chiral-trans}).  
By making use of the algebraic properties of the
$\bm{\tau}$ matrices, one can find an expression for the matrix $h(x)$,
which determines axial transformations of the fields,
to lowest order in $\mbox{\boldmath $\alpha$}$ but to {\em all\/} 
orders in the pion field:
\begin{equation}
{\gamma}^a = - {\epsilon}^{abc} \: {\alpha}^b \: \hat{\pi}^c \: {\rm tan} 
\left( { {\pi}\over {2 f_{\pi}} } \right) + O({\alpha}^2) \ .
\end{equation}
Note that $\pi \equiv |\mbox{\boldmath $\pi$}|$ {\em everywhere\/} in this 
paper (i.e., $\pi$ is {\em never\/} a shorthand for  3.14159\dots), 
$\hat{\pi} \equiv \mbox{\boldmath $\pi$}/\pi$, 
and we use Latin indices 
$a, b, c, \dots, \ell , m, \dots\ $ to denote isospin components.
When expanded to lowest order in the pion field, this result produces a
familiar expression~\cite{Fur97}:
\begin{equation}
{\gamma}^a = - { {1}\over {2 f_{\pi}} } \, [\mbox{\boldmath $\alpha
   \times \pi$}]^a + O({\alpha}^2, {\pi}^3) \ .
\end{equation}

We turn now to the ${\cal L}_{\pi}$ part of the lagrangian, which contains
only pion fields.
By writing the $U$ matrix as
\begin{equation}
U = {\rm exp} \left({ {i \over {f_{\pi}} } \,
{\mbox{\boldmath $ \tau  \cdot \pi$}} } \right) 
= {\rm cos} \left( { {\pi} \over {
f_{\pi} } } \right) + i {\mbox{\boldmath $ \tau \cdot \pi$}}
 \: {1 \over {\pi} } \: {\rm sin} \left( { {\pi} \over {f_{\pi} } } 
\right) \ ,
\label{eq:u-matrix}
\end{equation}
one can calculate the canonical momentum conjugate to the 
pion field to all orders in $\pi$:
\begin{equation}
P^a_{\pi} \equiv { {\partial {\cal L}_{\pi}} \over {\partial ({\partial}_0
{\pi}^a) } } = \left[ A_1^2 (\pi ) \left({\delta}^{ab} - \hat{\pi}^a
\hat{\pi}^b \right) + \hat{\pi}^a \hat{\pi}^b \right] \, {\partial}_0
{\pi}^b \ ,
\label{eq:pi-2-momentum}
\end{equation}
where
\begin{equation}
A_1(\pi) \equiv \left[ {{f_{\pi}} \over {\pi} } \: {\rm sin} \left( {
{\pi} \over {f_{\pi}} } \right) \right] \ .
\label{eq:A_1}
\end{equation}
Equation (\ref{eq:pi-2-momentum})
can be inverted to obtain ${\partial}_0 {\pi}^a$ in terms of
the canonical momentum $P^a_{\pi}$, 
which is required for evaluating the charge algebra:
\begin{equation}
{\partial}_0 {\pi}^a = \left[ { 1 \over {A_1^2} (\pi) } 
\left({\delta}^{ab} -
\hat{\pi}^a \hat{\pi}^b \right) + \hat{\pi}^a \hat{\pi}^b \right]
P^b_{\pi} \ .
\label{eq:pi-2-inverse}
\end{equation}

Next one can write out Noether currents corresponding to the
${\cal L}_\pi$ lagrangian in terms of the $U$ matrices. 
Here Noether
currents are defined according to \cite{IZ}
\begin{equation}
{\cal J}^{a \mu} \equiv - { {\partial  {\cal L} (\phi',
\partial \phi') } \over {\partial \left({\partial}_{\mu}
{\epsilon}^a (x) \right) } } \ ,
\label{eq:noether-definition}
\end{equation}
where $\epsilon^a (x)$ is a set of local, 
infinitesimal transformation parameters.

The lagrangian (\ref{eq:lagrangian-pi}) produces the following
currents: a vector current
\begin{equation}
V_{\pi}^{a \mu} = - i\,{ {f_{\pi}^2} \over 4} \, {\rm Tr} 
\left\{ {\tau}^a \left( U {\partial}^{\mu} U^{\dag} + U^{\dag}
{\partial}^{\mu} U \right) \right\}
\label{eq:vector-current}
\end{equation}
and an axial-vector current
\begin{equation}
A_{\pi}^{a \mu} = - i\,{ {f_{\pi}^2} \over 4}\, {\rm Tr} 
\left\{ {\tau}^a \left( U {\partial}^{\mu} U^{\dag} - U^{\dag}
{\partial}^{\mu} U \right) \right\} \ .
\label{eq:axial-current}
\end{equation}
[These are both conserved if the pion mass is zero.
For finite pion mass, one can derive PCAC from 
Eq.~(\ref{eq:noether-definition}).]
The corresponding Noether charge densities can be calculated 
in terms of the pion canonical momentum by substituting the
expression (\ref{eq:u-matrix}) into the general expressions for 
the currents (\ref{eq:vector-current}) and (\ref{eq:axial-current}),
and by rewriting the time derivative of
the pion field in terms of the canonical momentum
according to Eq.~(\ref{eq:pi-2-inverse}).  
One obtains, for the vector charge density,
\begin{equation}
V^{a0}_{\pi} = A_1^2 ( \pi ) \: [ \mbox{\boldmath$\pi \times {}$}  
{\partial}_0 
\mbox{\boldmath $\pi$} ]^a = 
[ \mbox{\boldmath$\pi \times P$}_{\pi} ]^a \ .
\label{eq:pi-2-vector-charge}
\end{equation}
Note that when written in terms of the canonical momentum, the
expression for the vector charge is identical to that in a linear 
representation of the chiral symmetry \cite{Ser86,Ser92}.
The expression for the axial charge density is more complicated:
\begin{equation}
A^{a0}_{\pi} = - \left[ B_0(\pi) \left({\delta}^{ab} - \hat{\pi}^a
\hat{\pi}^b \right) + f_{\pi} \, \hat{\pi}^a \hat{\pi}^b \right]
P^b_{\pi} \ ,
\label{eq:pi-2-axial-charge}
\end{equation}
where
\begin{equation}
B_0(\pi) \equiv \pi \; {\rm cotan} \left( { {\pi} \over f_{\pi} } 
\right) \ .
\label{eq:B_0}
\end{equation}

By utilizing the usual boson commutator to quantize the pion field:
\begin{equation}
\left[ P^b_{\pi} ({\mbox{\boldmath $x$}}, t), {\pi}^a
({\mbox{\boldmath $y$}}, t) \right] = -i {\delta}^{ab} 
\delta^{(3)}
({\mbox{\boldmath $x$}} -{\mbox{\boldmath $y$}} ) \ ,
\label{eq:pi-commutator}
\end{equation}
and by calculating the commutators of the vector and axial charges
\begin{equation}
Q^a \equiv \int\! d^{\, 3} x \:  V^{a0} \ , \quad
Q^a_5 \equiv \int\! d^{\, 3} x \:  A^{a0} \ , 
\end{equation}
one finds, after some algebra, the familiar results:
%
%\begin{equation}
\beqa
\left[ Q^a, Q^b \right] & = & i {\epsilon}^{abc} \: Q^c \ ,
   \label{eq:chiral-algebra-vv}
    \\[6pt]
%\end{equation}
%
%\begin{equation}
\left[ Q^a, Q^b_5 \right] & = & i {\epsilon}^{abc} \: Q^c_5 \ ,
   \label{eq:chiral-algebra-va}
    \\[6pt]
%\end{equation}
%
%\begin{equation}
\left[ Q^a_5, Q^b_5 \right] & = & i {\epsilon}^{abc} \: Q^c \ .
   \label{eq:chiral-algebra-aa}
%\end{equation}
\eeqa

It is interesting to note that the explicit form of 
$B_0 (\pi )$ is needed
to prove only the last relation (\ref{eq:chiral-algebra-aa}). 
The evaluation of the vector charge commutator
(\ref{eq:chiral-algebra-vv}) is identical to that in the linear
theory, while the second relation (\ref{eq:chiral-algebra-va})
illustrates that the axial-vector charge is an isovector. 
Both of these results hold because the $SU(2)_V$
symmetry is represented linearly in the present approach.

\subsection{Pion--nucleon terms with $\nu = 2$}

Let us now include nucleons in the analysis. At order $\nu =2$ in the
NDA counting scheme, the additional piece in the lagrangian 
is given in Eq.~(\ref{eq:lagrangian-pi-N}).  
The corresponding extra piece in the pion canonical momentum is
\begin{equation}
{ {\partial {\cal L}_{\pi N}} \over {\partial ({\partial}_0 {\pi}^a) } } =
{\epsilon}^{abc} \: {\pi}^b \: {1 \over {\pi}^2 } \:
{\rm sin}^2 \left( { {\pi} \over {2 f_{\pi}} } \right) N^{\dag} {\tau}^c N
+ { 1 \over {2 f_{\pi} } } \: g_A \left[  A_1(\pi) \left({\delta}^{ab} - 
\hat{\pi}^a \hat{\pi}^b \right) + \hat{\pi}^a \hat{\pi}^b \right] 
N^{\dag} {\gamma}_5 {\tau}^b N \ ,
\label{eq:mometum-pi-N}
\end{equation}
where $A_1(\pi)$ is defined in Eq.~(\ref{eq:A_1}).
(The nucleon canonical momentum is $i N^\dagger$, as usual.)
The full pion canonical momentum becomes
\begin{equation}
P^a_{\pi} \equiv { {\partial {\cal L}_{\pi}} \over {\partial ({\partial}_0
{\pi}^a) } } + { {\partial {\cal L}_{\pi N}} \over {\partial
({\partial}_0 {\pi}^a) } } \ ,
\label{eq:nu2-momentum}
\end{equation}
and one can invert this relation using the same projector as
before. [See Eq.~(\ref{eq:pi-2-inverse}).]
The resulting expression for ${\partial}_0 {\pi}^a$ is
\begin{eqnarray}
{\partial}_0 {\pi}^a & = & \left[ { 1 \over {A_1^2} ( \pi ) }
 \left({\delta}^{ab} - \hat{\pi}^a \hat{\pi}^b \right) + \hat{\pi}^a
 \hat{\pi}^b \right] P^b_{\pi} 
 - {\epsilon}^{abc} {1 \over {A_1^2} (\pi )} \,
 {\pi}^b {1 \over {{\pi}^2} } \: {\rm sin}^2 \left( { {\pi} \over {2
 f_{\pi}} } \right) N^{\dag} {\tau}^c N \nonumber \\[5pt]
 &  & \quad - { 1
 \over {2 f_{\pi} }}\: 
 g_A \left[ {1 \over{ A_1} ( \pi )} \left( {\delta}^{ab}
 - \hat{\pi}^a \hat{\pi}^b \right) + \hat{\pi}^a \hat{\pi}^b \right]
 N^{\dag} {\gamma}_5 {\tau}^b N \ .
\label{eq:pi-N-momentum-inverse}
\end{eqnarray}
We will see that the substitution of this expression into the relations
for the charges makes the latter look simple.

The $\pi$N contributions to the Noether currents can be written
in terms of the $\xi$ matrix as
\begin{equation}
V^{a \mu}_{\pi N} = { {1} \over {4} }\, \Nbar {\gamma}^{\mu} 
\left[ {\xi} {\tau}^a  {\xi}^{\dag} + {\xi}^{\dag} {\tau}^a 
 {\xi} \right] N
+ { {1} \over {4} }\, g_A \Nbar {\gamma}^{\mu} {\gamma}_5
\left[ {\xi} {\tau}^a  {\xi}^{\dag} - {\xi}^{\dag} {\tau}^a 
 {\xi} \right] N \ ,
    \label{eq:vector-currentpiN}
\end{equation}
\begin{equation}
A^{a \mu}_{\pi N} = - { {1} \over {4} } \,\Nbar {\gamma}^{\mu} 
\left[ {\xi} {\tau}^a  {\xi}^{\dag} - {\xi}^{\dag} {\tau}^a 
 {\xi} \right] N
- { {1} \over {4} }\, g_A \Nbar {\gamma}^{\mu} {\gamma}_5
\left[ {\xi} {\tau}^a {\xi}^{\dag} + {\xi}^{\dag} {\tau}^a 
 {\xi} \right] N \ ,
    \label{eq:axial-currentpiN}
\end{equation}
which include the pion field to all orders.
By substituting an expression for $\xi$ analogous to 
Eq.~(\ref{eq:u-matrix}) and by performing some algebra,
we find
\begin{equation}
V^{a \mu}_{\pi N} = \Nbar {\gamma}^{\mu} { {{\tau}^b} \over 2 } N
\left[ {\rm cos} \left( { {\pi} \over {f_{\pi} } } \right)
({\delta}^{ab} - \hat{\pi}^a \hat{\pi}^b) + \hat{\pi}^a \hat{\pi}^b
\right] + g_A \: {\epsilon}^{abc} \: \hat{\pi}^b \: {\rm sin} \left( {
{\pi} \over {f_{\pi} } } \right) \: \Nbar {\gamma}^{\mu} {\gamma}_5
{ {{\tau}^c} \over 2} N \ ,
\label{eq:V-pi-N}
\end{equation}
\begin{equation}
A^{a \mu}_{\pi N} = - {\epsilon}^{abc} \: \hat{\pi}^b \: {\rm sin}
\left( { {\pi} \over {f_{\pi} } } \right) \: \Nbar {\gamma}^{\mu} {
{{\tau}^c} \over 2} N - g_A \Nbar {\gamma}^{\mu} {\gamma}_5 {
{{\tau}^b} \over 2 } N \left[ {\rm cos} \left( { {\pi} \over {f_{\pi}
} } \right) \: \left({\delta}^{ab} - \hat{\pi}^a \hat{\pi}^b \right) +
\hat{\pi}^a \hat{\pi}^b \right] \ .
\label{eq:A-pi-N}
\end{equation}
One can combine these results with those in
Eqs.~(\ref{eq:vector-current}) and (\ref{eq:axial-current})
to construct the currents
\begin{equation}
V^{a \mu}_{2} \equiv V^{a \mu}_{\pi} + V^{a \mu}_{\pi N} \ ,
\label{eq:V-2}
\end{equation}
\begin{equation}
A^{a \mu}_{2} \equiv A^{a \mu}_{\pi} + A^{a \mu}_{\pi N} \ .
\label{eq:A-2}
\end{equation}

By utilizing the new expression for the pion canonical momentum,
including the pion--nucleon interaction contributions, 
the corresponding charge densities can be expressed in terms of
canonical momenta. 
The vector charge density is again precisely as in
the linear model:
\begin{equation}
V^{a0}_{2} = [ \mbox{\boldmath$\pi \times P$}_{\pi} ]^a
 + N^{\dag} { {{\tau}^a} \over 2 } N \ .
\label{eq:charge-2-vector}
\end{equation}
The expression for the axial charge density is a bit more complicated:
\begin{equation}
A^{a0}_{2} = - \left[ B_0(\pi) \left({\delta}^{ab} - 
\hat{\pi}^a \hat{\pi}^b \right) 
  + f_{\pi} \, \hat{\pi}^a \hat{\pi}^b \right]
P^b_{\pi} - B_1 (\pi) \: {\epsilon}^{abc} \: \hat{\pi}^b \: N^{\dag} {
{{\tau}^c} \over 2} N \ ,
\label{eq:charge-2-axial}
\end{equation}
where $B_0(\pi)$ is defined in Eq. (\ref{eq:B_0}) and
\begin{equation}
B_1(\pi) \equiv {\rm tan} \left(  { {\pi} \over {2 f_{\pi}} } 
\right) \ .
\label{eq:B_1}
\end{equation}

Canonical quantization is carried out using the commutator
(\ref{eq:pi-commutator}) for the pion and the anticommutator 
\begin{equation}
\left\{ N_{\alpha} ({\mbox{\boldmath $x$}}, t), N^{\dag}_{\beta} \,
 ({\mbox{\boldmath $y$}}, t) \right\} = {\delta}_{\alpha \beta} \:
 \delta^{(3)} ({\mbox{\boldmath $x$}} -{\mbox{\boldmath $y$}} )
\label{eq:N-commutator}
\end{equation}
for the nucleon fields.
The conserved charges are defined as above, and
after some algebra, one can prove that the same correct chiral
charge algebra given in Eqs.~(\ref{eq:chiral-algebra-vv}) 
to (\ref{eq:chiral-algebra-aa}) holds. 
We note that, as before, only the proof of the commutator of
two axial charges requires the explicit forms of 
$B_0 (\pi )$ and $B_1 (\pi )$.

\section{Interaction amplitudes}
\label{sec:amps}

Here we consider interaction amplitudes originating from the $\nu =2$
terms in the lagrangian of Eq.~(\ref{eq:lagrangian-nu2}).  
One can write out
the interaction vertices resulting from this lagrangian 
that are important for calculating one- and two-body current matrix
elements and pion-production amplitudes.

\begin{figure}[htb]
\bigskip
\begin{center}
\includegraphics*[width=1.85in,angle=0]{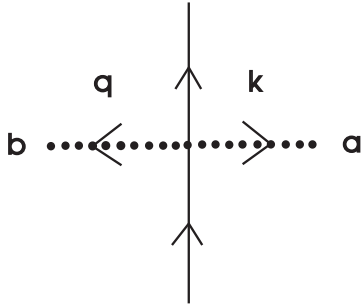}

  \caption{The pion--pion--nucleon vertex from 
           Eq.~(\protect\ref{eq:piNvertices}).
           Here $q^\mu \equiv p^{\mu}_i - p^{\mu}_f - k^{\mu}$, in
           terms of the initial and final nucleon four-momenta
           and the outgoing pion four-momentum $k^{\mu}$.}
      \label{fig:pi-pi-n0}
\end{center}
\end{figure}

\begin{figure}[htb]
\bigskip
\begin{center}
\includegraphics*[width=1.02in,angle=0]{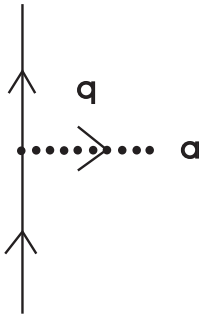}

 \caption{The pion--nucleon vertex from
          Eq.~(\protect\ref{eq:piNvertices}).
          Here $q^\mu \equiv p^{\mu}_i - p^{\mu}_f$.}
     \label{fig:pi-n}
\end{center}
\end{figure}

The lowest-order strong-interaction $\pi$N vertices originate from
the following interaction terms in the lagrangian 
(\ref{eq:lagrangian-pi-N}):
\begin{equation}
{\cal L}_{\rm int} \approx - { 1 \over {4 f_{\pi}^2} } \,{\epsilon}^{abc}
{\pi}^a \Nbar (\not \! \partial {\pi}^b) {\tau}^c N + { g_A \over {2
f_{\pi}} } \,\Nbar (\not \! \partial {\pi}^a) {\gamma}_5 {\tau}^a N \ .
\end{equation}
The analytic forms for these vertices are
\begin{equation}
\frac{1}{4 f_\pi^2} \, \epsilon^{abc} \not \! q \,\tau^c \ , \qquad
\frac{g_A}{2f_\pi} \, \gamma_5 \not \! q \,\tau^a \ ,
   \label{eq:piNvertices}
\end{equation}
respectively; these are to be used in the Feynman rules for the calculation
of the $S$ matrix~\cite{Ser86,IZ,MS}.
Diagrammatically, these interaction vertices can be represented as in
Figs.~\ref{fig:pi-pi-n0} and \ref{fig:pi-n},
where the solid lines denote the nucleons and
the dotted lines are the pions.

%%%%%%%%%%%%%%%%%%%%%%%%%%%%%%%%%%%%%%%%%%%%%%%%%%%%%%%%
%
%%%%%%%%%%%%%%%%%%%%%%%%%%%%%%%%%%%%%%%%%%%%%%%%%%%%%%%%

To lowest order in the pion field, the Noether axial current due
to ${\cal L}_2$ takes the form
\begin{equation}
A^{a \mu}_2 = - {1 \over f_{\pi}} \, \epsilon^{abc} \pi^b \,
\Nbar \gamma^\mu {\tau^c \over 2}\, N
- g_A \, \Nbar \gamma^\mu \gamma_5 {\tau^a \over 2}\, N
- f_{\pi} \, {\partial}^{\mu} {\pi}^a + O({\pi}^2) \ .
    \label{eq:firstaxial}
\end{equation}
To determine the Feynman rules for the axial-current vertices, we consider
a lagrangian density
\begin{equation}
 {\cal L}_{\mathrm{ext}} (x) = 
          A^{a \mu} (x) \, S^{\mathrm{ext}}_{a \mu} (x) \ ,
  \label{eq:Lext}
\end{equation}
where $S^{\mathrm{ext}}_{a \mu}$ is an external source that could originate
from leptons, for example.
The scattering matrix $S_{fi}$ can then be written to first order in
the external source as
\begin{equation}
 S^{\mkern2mu (1)}_{fi} = \left( \:
   \prod_{\mathrm{bosons}} {1 \over (2 \omega_b {\mathbb V})^{1/2}}
   \right)  \left( \:
\prod_{\mathrm{fermions}} {1 \over (2 E_f {\mathbb V})^{1/2} } \right)
 M^{a \mu} (k) \, {\widetilde S}^{\mathrm{ext}}_{a \mu} (k) \ ,
    \label{eq:Sext}
\end{equation}
which {\em defines} the covariant amplitude $M^{a \mu} (k)$.
Here $\omega_b$ are the boson energies, $E_f$ are the fermion energies,
and $\mathbb V$ is the quantization volume; these factors
specify the normalization to be used on external lines.
The fermion spinors are
included in $M^{a \mu}$, which
is to be computed using the Feynman rules with
the axial-current vertices given below; we assume overall four-momentum
conservation (as well as four-momentum conservation at every vertex), 
and we adopt the covariant spinor
normalization $\bar{u} (p) u(p) = -\bar{v} (p) v(p) = 2M$.

%%%%%%%%%%%%%%%%%%%%%%%%%%%%%%%%%%%%%%%%%%%%%%%%%%%%%%%%
\begin{figure}[htb]
%\bigskip
\medskip
\begin{center}
\includegraphics*[width=1.4in,angle=0]{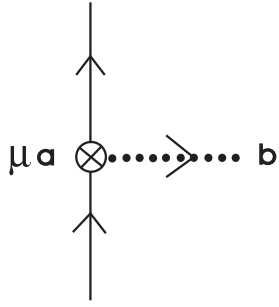}

  \caption{The nucleon--pion vertex in the axial current
           (\protect\ref{eq:firstaxial}), from
           Eq.~(\protect\ref{eq:axialvertices}).
           The crossed circle shows where and how to attach
           the external source 
           ${\widetilde S}^{\mathrm{ext}}_{a \mu} (k)$.}
      \label{fig:pi-a-n}
\end{center}
\end{figure}
\begin{figure}[htb]
%\bigskip
\medskip
\begin{center}
\includegraphics*[width=0.5in,angle=0]{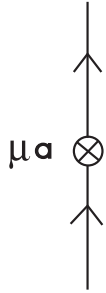}

  \caption{The vertex for the nucleon-only term in
           the axial current (\protect\ref{eq:firstaxial}),
           from Eq.~(\protect\ref{eq:axialvertices}).}
      \label{fig:a-n}
\end{center}
\end{figure}
\begin{figure}[htb]
\bigskip
\begin{center}
\includegraphics*[width=1.4in,angle=0]{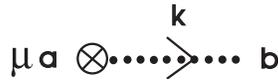}

  \caption{The vertex for the pionic
           contribution to the axial current
           (\protect\ref{eq:firstaxial}), from
           Eq.~(\protect\ref{eq:axialvertices}).}
      \label{fig:pi-a}
\end{center}
\end{figure}
%
%%%%%%%%%%%%%%%%%%%%%%%%%%%%%%%%%%%%%%%%%%%%%%%%%%%%%%%%

The vertices arising from Eqs.~(\ref{eq:firstaxial}) to
(\ref{eq:Sext}) are given by
\begin{equation}
- {i \over f_{\pi}} \, \epsilon^{abc} \gamma^\mu {\tau^c \over 2}
 \ , \qquad
- i {g_A}\mkern2mu
  \gamma^\mu \gamma_5 {\tau^a \over 2}\ , \qquad
  f_{\pi} k^{\mu} {\delta}^{ab} \ ,
     \label{eq:axialvertices}
\end{equation}
and they are represented diagrammatically in
Figs.~\ref{fig:pi-a-n}, \ref{fig:a-n}, and \ref{fig:pi-a},
respectively.

%%%%%%%%%%%%%%%%%%%%%%%%%%%%%%%%%%%%%%%%%%%%%%%%%%%%%%%%
\begin{figure}[htb]
%\bigskip
\medskip
\begin{center}
\includegraphics*[width=3.42in,angle=0]{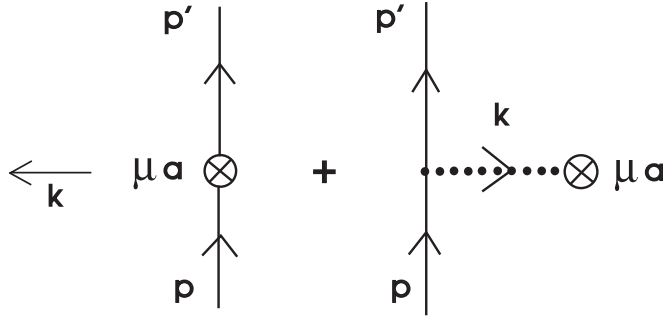}

  \caption{The leading-order, one-body, axial-vector current.
           Note that the external source 
           ${\widetilde S}^{\mathrm{ext}}_{a \mu} (k)$
           ``extracts'' the momentum $k^{\mu}$ from {\em both\/}
           diagrams: {\em explicitly\/} when the pion leaves the
           diagram and {\em implicitly\/} when the source attaches
           directly to the nucleon line.}
      \label{fig:A-1N}
\end{center}
\end{figure}
%
%%%%%%%%%%%%%%%%%%%%%%%%%%%%%%%%%%%%%%%%%%%%%%%%%%%%%%%%

The one-body, axial-current matrix element to lowest order in
$1/f_{\pi}$ is given by
\begin{equation}
 M^{a \mu} (1)
= i g_A \: \bar{u}(p') {\gamma}_5 \left\{ {\gamma}^{\mu} -
\rlap/{\mkern-1mu k} 
{ {k^{\mu}} \over {k^2-m^2_{\pi}} } \right\} { {{\tau}^a}
\over 2 }\, u(p) \ .
    \label{eq:axial-current-nucleon}
\end{equation}
The two relevant Feynman diagrams are drawn
in Fig.~\ref{fig:A-1N},
where $k^{\mu}$ is the {\em outgoing\/} momentum 
on the pion line.
Note that the multiplicative factor of $g_A$ 
in the amplitude implies that
{\em no additional current renormalization is required.}
Here $g_A$ enters as an overall factor, which was missing in the 
tree-diagram amplitude calculated in the $\sigma\omega$ model of
Ref.~\cite{Ana98}:
\begin{equation}
 M^{a \mu} (1) = g_A \: M^{a \mu}_{\sigma\omega} (1) \ .
\end{equation}
The interaction amplitude (\ref{eq:axial-current-nucleon}) satisfies
PCAC automatically, due to the presence of the projection operator in
braces:
\begin{equation}
k_{\mu} M^{a \mu} (1)
%%%%%%%  \Inthelimit
\lower1.9ex\vbox{\hbox{$\  
       \buildrel{\hbox{\Large \rightarrowfill}}
       \over{\scriptstyle{m_{\pi} \, \rightarrow \, 0}}\ $}}
\, 0 \ .
\end{equation}

%%%%%%%%%%%%%%%%%%%%%%%%%%%%%%%%%%%%%%%%%%%%%%%%%%%%%%%%
\begin{figure}[htb]
\bigskip\bigskip
\begin{center}
\includegraphics*[width=5.5in,angle=0]{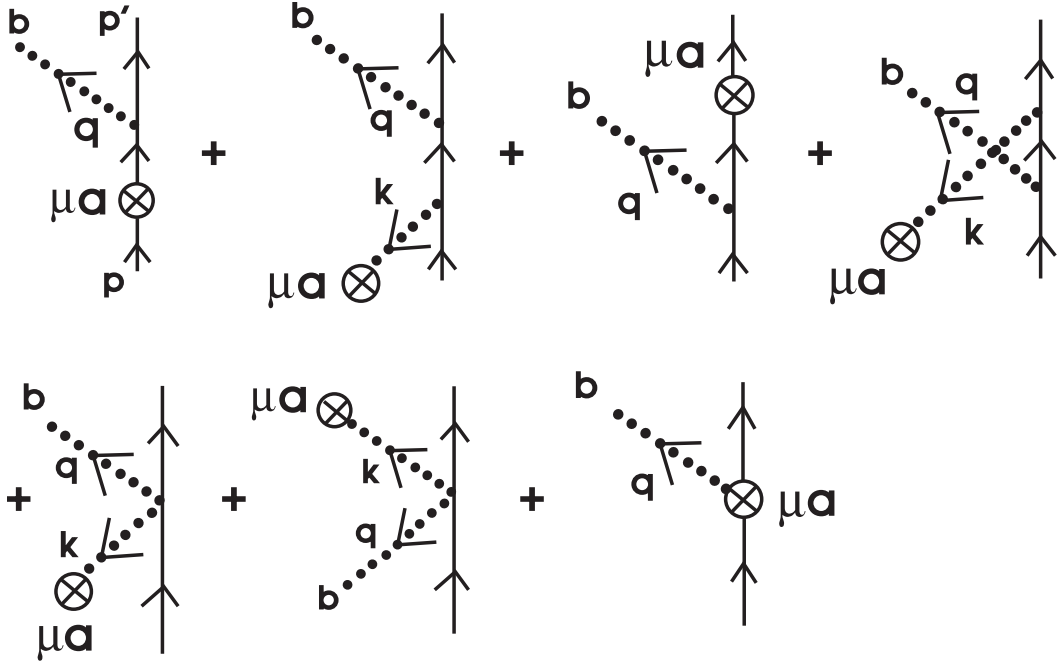}

   \caption{The amplitude for axial-current pion production 
            on a single nucleon [Eq.~(\protect\ref{eq:piprod-2})].
            Again, note that $k^{\mu}$ is extracted by the external
            source, so that $p^{\mu} = {p'}^{\mu} + k^{\mu}
            + q^{\mu}$.}
       \label{fig:pi-pi-n}
\end{center}
\end{figure}
%
%%%%%%%%%%%%%%%%%%%%%%%%%%%%%%%%%%%%%%%%%%%%%%%%%%%%%%%%

The amplitude for axial-current pion production on a
single nucleon is represented by the diagrams of
Fig.~\ref{fig:pi-pi-n} and can be written as
\begin{eqnarray}
 2 f_{\pi} \: M^{ab\, \mu} ( \pi ) & = & 
 \bar{u}(p') \left\{ g^2_A \left[ \not \! q \, {\gamma}_5 \, {1 \over
 {(\not \! p^{\mkern2mu \prime} \, 
 + \not \!  q)-M}} \, {\gamma}_5 \left( {\gamma}^{\mu} -
 \not \! k { {{k}^{\mu}} \over {k^2 - m_{\pi}^2} } \right) {
 { {\tau}^b {\tau}^a } \over 2} \right. \right. \nonumber \\[6pt] 
&  & \qquad\qquad\qquad\quad +
 \left. \left( {\gamma}^{\mu} - \not \! {k} { {k^{\mu}} \over {k^2 -
 m_{\pi}^2} } \right) {\gamma}_5 {1 \over {(\not \! p \; -
\not \! q) - M}} \,{\gamma}_5 \!\not \! q \,
 { { {\tau}^a {\tau}^b } \over 2} \right] \nonumber \\[6pt] 
&  & \qquad\qquad
 + \left. i {\epsilon}^{abc} \, { {{\tau}^c} \over 2} \left[ (\not
 \! k\, - \!\not \! q) { {k^{\mu}} \over {k^2 - m_{\pi}^2}} - 2
 {\gamma}^{\mu} \right] \right\} u(p) \ ,
   \label{eq:piprod-2}
\end{eqnarray}
where $q^{\mu}$ is the outgoing four-momentum of the emitted pion,
and $p^{\mu}$ and ${p'}^{\mu}$ are the initial and final
nucleon four-momenta, respectively.
It is easy to verify that this amplitude satisfies PCAC 
(for an on-shell nucleon).

In the soft-pion limit $( q, k \rightarrow 0)$ and to leading order
in $1/M$,
\begin{equation}
{1 \over { {\mathbb V} {(2 E \, 2 E')}^{1/2} }} 
\: iM^{ab\, \mu} (\pi ) \longrightarrow {1 \over
{f_{\pi}} } \: {\epsilon}^{abc} \, {{\tau}^{c} \over 2} \: 
{\delta}^{\mu 0} \ ,
\end{equation}
where $E \approx E' \approx M$ are the initial and
final energies of the nucleons, and the arrow signifies that we
have removed the external nonrelativistic wave functions
(including a factor of $1/\sqrt{\mathbb V}\,$) to arrive at a
first-quantized operator.
This amplitude reproduces the result of Kubodera, Delorme, and Rho
(KDR)~\cite{Kub78} obtained in the current-algebra approach and solves
the problem of an extra factor of $g^2_A$ found in Ref.~\cite{Ana98}.

\begin{figure}[htbp]
\bigskip\medskip
\begin{center}
\includegraphics*[width=3.4in,angle=0]{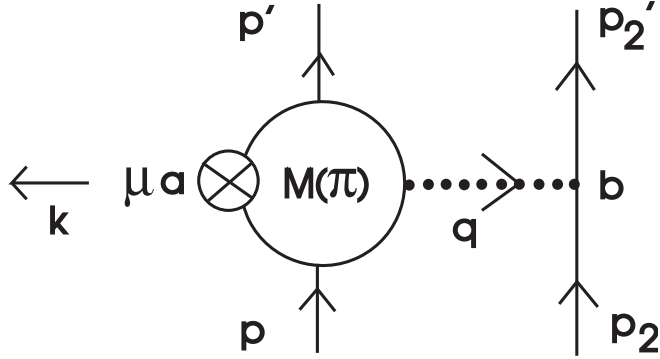}

  \caption{Two-nucleon contributions to the axial current.
        Here the vertex $M(\pi )$ 
        [Eq.~(\protect\ref{eq:piprod-2})]
        represents all possible
        ways that the external source can couple to the
        left-hand nucleon line (to the order we are working),
        as illustrated in Fig.~\protect\ref{fig:pi-pi-n}.}
      \label{fig:pi-a-2n0}
\end{center}
\end{figure}

Next we consider amplitudes for the axial current that involve two 
nucleons.
The relevant diagrams to this order in $\nu$ are presented in
Fig.~\ref{fig:pi-a-2n0} (with a similar set of diagrams when the 
external source interacts with the second nucleon) and produce an
amplitude of the form
\begin{equation}
  M^{a \mu} (2) =  
    M^{ab \: \mu} (\pi)\; { i \over {q^2 - m_{\pi}^2}}
 \; \, { {g_A} \over {f_{\pi}} } \; 
 \bar{u}(p'_2) \not \! q \,{\gamma}_5 \,
 { {{\tau}^b} \over 2} \, u(p_2) + \mathrm{(direct)}_2
  + \mathrm{cross\ terms.}
   \label{eq:L2-axial-2N-amplitude}
\end{equation}
Here ``$\mathrm{(direct)}_2$'' denotes the terms where the 
external source hits nucleon 2, and
``cross terms'' denotes similar diagrams with the fermion lines
crossed in the final state.
%
%%%%%%%%%%%%%%%%%%%%%%%%%%%%%%%%%%%%%%%%%%%%%%%%%%%%%%%%
%
%%%%%%%%%%%%%%%%%%%%%%%%%%%%%%%%%%%%%%%%%%%%%%%%%%%%%%%%
%
This amplitude satisfies PCAC because $M^{ab \: \mu} (\pi)$ does. 
The corresponding
two-body nuclear AXC can be identified to lowest order in the
inverse nucleon mass $M$ from an expression for the 
$S$-matrix element:\footnote{%
In this paper, we do not discuss the details of the derivation
(from the covariant amplitudes) of the axial-vector exchange
currents that are to be used with 
relativistic, mean-field, four-component Dirac wave 
functions~\protect\cite{Fur97}.
For some of the issues that are involved,
see, for example, Refs.~\protect\cite{Wal95,Dub76,Dmi98,Ana98}.}
\begin{equation}
   {A}_{\mathrm{eff}}^{a \mu}(2) \equiv
 - {1 \over {{(2M \, {\mathbb V} )}^2}} \, iM^{a \mu} (2) \ .
  \label{eq:twob-id}
\end{equation}
By performing a nonrelativistic reduction of the direct terms
in the resulting current, and by removing factors for the external
nonrelativistic wave functions,
one obtains\footnote{%
The ``cross terms'' will be included when one takes matrix 
elements of this first-quantized, two-body operator between
antisymmetrized many-fermion wave functions.}
\begin{equation}
  A_{\mathrm{eff}}^{a0}({\bf q},{\bf Q}) 
      \longrightarrow  {g_A \over {4 f_{\pi}^2} } \;
i \, {\left[ \mbox{\boldmath$\tau$}(1) 
 \mbox{\boldmath${} \times \tau$}(2) \right]}^{a}\;
\left\{
{ {\mbox{\boldmath
$\sigma$}}(2) \mbox{\boldmath$\,\cdot\,$}
 {{\bf q}} \over {{{\bf q}}^2 + m_{\pi}^2} } - 
{ {\mbox{\boldmath
$\sigma$}}(1) \mbox{\boldmath$\,\cdot\,$}
 {{\bf Q}} \over {{{\bf Q}}^2 + m_{\pi}^2} }
\right\} \ ,
\label{eq:nu2-axial-charge}
\end{equation}
where the exchanged momenta are defined by
${\mathbf q}  \equiv  {\mathbf p}'_2 - {\mathbf p}_2$ and
${\mathbf Q}  \equiv  {\mathbf p}' - {\mathbf p}$.

This is the same two-body, axial charge density obtained by
KDR~\cite{Kub78} in the current-algebra approach. 
Here, however, this result has been calculated as the lowest-order 
approximation to an EFT lagrangian.
The lagrangian approach allows us to
explicitly calculate additional contributions to this AXC from the
next-to-leading-order ($\nu =3$) terms in the NDA counting scheme.

\section{Relevant Terms with $\ \nu = 3$}
\label{sec:nuthree}

We now concentrate on the contributions from the $\nu = 3$
terms in the lagrangian. 
Since our ultimate objective is to identify 
axial-vector exchange currents arising from
Eq.~(\ref{eq:eft-lagrangian}) 
to lowest order in the meson fields, we consider only terms 
that contribute to the vector and axial-vector
currents to this order in $\nu$.
These terms are bilinear in derivatives of the pion field itself, but
contain the field to all orders.
We obtain the $\nu = 3$ lagrangian\footnote{%
We omit a $\nu = 3$ term with an antisymmetrized derivative on the
nucleon fields~\protect\cite{Ell98}
because it is effectively of order $\nu = 4$.} 

\begin{equation}
{\cal L}_{3} = {\delta}_1
{\cal L} + {\delta}_2 {\cal L}  \ ,
\label{eq:L3}
\end{equation}
where
\begin{equation}
{\delta}_1 {\cal L} \equiv - { { {\kappa}_{\pi} } \over M } 
\, \Nbar {v}_{\mu \nu} {\sigma}^{\mu \nu} N 
\label{eq:delta-1-lagrangian}
\end{equation}
and
\begin{equation}
{\delta}_2 {\cal L} \equiv { {4 {\beta}_{\pi}} \over M } \, \Nbar N
\, {\rm Tr} \left( a_{\mu} a^{\mu} \right) \ ,
\label{eq:delta-2-lagrangian}
\end{equation}
with $v_{\mu \nu}$ given by Eq.~(\ref{eq:v-mu-nu}), and
$a_{\mu}$ given by Eq.~(\ref{eq:a-mu}).
The final term contributing to the two-body
axial current at relevant order in $\nu$ involves the $\rho$ meson field:
\begin{eqnarray}
{\cal L}_{\rho} & \equiv &
   {-g_{\rho}} \Nbar {\rho}_{\mu} {\gamma}^{\mu} N 
-  { {f_{\rho} g_{\rho}} \over {4 M} } \,
\Nbar {\rho}_{\mu \nu} {\sigma}^{\mu \nu} N - { 1 \over 2 } \, 
{\rm Tr} \left( {\rho}_{\mu \nu} {\rho}^{\mu \nu} \right) 
\nonumber \\
 & & \quad
 {-g_{\rho \pi \pi}} \, { {2 f^2_{\pi}} \over { m^2_{\rho} } }
\, {\rm Tr} \left( {\rho}_{\mu \nu} {v}^{\mu \nu} \right)
+ m^2_{\rho} \, {\rm Tr} \left( {\rho}_{\mu} {\rho}^{\mu} 
\right) \ ,
\label{eq:delta-rho-lagrangian}
\end{eqnarray}
where $\rho_{\mu \nu}$ is defined in Eqs.~(\ref{eq:rhofieldtensor})
and (\ref{eq:rhoderiv}). 
Here the first and last terms are of order $\nu = 2$, the second
term is of order $\nu = 3$, and the third and fourth terms are
of order $\nu = 4$.
The fourth term is
included in ${\cal L}_{\rho}$ because it is the lowest-order
term that involves $\rho\pi$ interactions, which are thought to
be relevant in AXC originating from heavy mesons~\cite{Iva79}. 
We first consider the terms in Eqs.~(\ref{eq:delta-1-lagrangian}) 
and (\ref{eq:delta-2-lagrangian}), which describe
the interaction of nucleons with pions,
and then return to calculate the additional contributions due to
the $\rho$ meson field in Sec.~\ref{sec:rho}.

One can calculate the change in ${\cal L}_{3}$ 
under vector or axial-vector transformations. 
These quantities are required for calculating the
respective Noether currents according to the definition
(\ref{eq:noether-definition}):
\begin{equation}
\delta {\cal L}_3 = { {{2 i \kappa}_{\pi}} \over {M} } \,
 \Nbar \left[ \delta a_{\mu}, \: a_{\nu} \right]
{\sigma}^{\mu \nu} N + { {8 {\beta}_{\pi}} \over {M} }\, \Nbar N \: 
{\rm Tr} \left( \delta a_{\mu} \, a^{\mu} \right) + O \left( (\delta a)^2
 \right) \ ,
\label{eq:L3-change}
\end{equation}
where $\delta a_{\mu}$ is different for the (local) vector
and axial-vector transformations:
\begin{eqnarray}
\delta a_{\mu} & = &
 {1 \over 4} \left( {\xi}^{\dag} {\tau}^a \xi - 
\xi {\tau}^a {\xi}^{\dag} \right) {\partial}_{\mu} {\beta}^a
\qquad \mathrm{vector},
\label{eq:delta-a-vector} \\[6pt]
\delta a_{\mu}  & = &
 {1 \over 4} \left( {\xi}^{\dag} {\tau}^a \xi + 
\xi {\tau}^a {\xi}^{\dag} \right) {\partial}_{\mu} {\alpha}^a
\qquad \mathrm{axial\ vector}.
\label{eq:delta-a-axial}
\end{eqnarray}
These variations produce the following Noether currents, 
as defined in Eq.~(\ref{eq:noether-definition}):
\begin{equation}
V^{a \mu}_3 = - { {i {\kappa}_{\pi}} \over {2 M} } \: \Nbar
{\sigma}^{\mu \nu} \left[ {\xi}^{\dag} {\tau}^a \xi - \xi {\tau}^a
{\xi}^{\dag}, \: a_{\nu} \right] N - { {2 {\beta}_{\pi}} \over {M} }
\: \Nbar N \: {\rm Tr} \left[ ( {\xi}^{\dag} {\tau}^a \xi - \xi
{\tau}^a {\xi}^{\dag} ) \: a^{\mu} \right] \ ,
\label{eq:vector-current-general}
\end{equation}
\begin{equation}
A^{a \mu}_3 
   = - { {i {\kappa}_{\pi}} \over {2 M} } \: \Nbar {\sigma}^{\mu \nu}
\left[ {\xi}^{\dag} {\tau}^a \xi + \xi {\tau}^a {\xi}^{\dag}, 
\: a_{\nu} \right] N
- { {2 {\beta}_{\pi}} \over {M} } \: \Nbar N \: {\rm Tr}
\left[ ( {\xi}^{\dag} {\tau}^a \xi 
+ \xi {\tau}^a {\xi}^{\dag} ) \: a^{\mu} \right] \ .
\label{eq:axial-current-general}
\end{equation}
For future reference, we exhibit some often-encountered expressions
as functions of the pion field:
\begin{equation}
\left( {\xi}^{\dag} {\tau}^a \xi + \xi {\tau}^a {\xi}^{\dag} \right) =
2 {\tau}^b 
\left( {\rm cos} \left( { {\pi} \over {f_{\pi}} } \right)  
\left[ {\delta}^{ab} - 
\hat{\pi}^a \hat{\pi}^b  \right] +
\hat{\pi}^a \hat{\pi}^b \right) \ ,
\label{eq:plus}
\end{equation}
\begin{equation}
\left( {\xi}^{\dag} {\tau}^a \xi - \xi {\tau}^a {\xi}^{\dag} \right) =
- 2 {\epsilon}^{abc} \hat{\pi}^b {\tau}^c \: {\rm sin} \left( { {\pi}
\over {f_{\pi}} } \right) \ ,
\label{eq:minus}
\end{equation}
\begin{equation}
a_{\mu} = { {1} \over {2 f_{\pi}} } \, 
  {\tau}^c \,{\partial}_{\mu} {\pi}^d
\left\{ A_1( \pi ) \left[ {\delta}^{cd} - \hat{\pi}^c \hat{\pi}^d
\right] + \hat{\pi}^c \hat{\pi}^d \right\} \ ,
\label{eq:a-mu-allorder}
\end{equation}
where $A_1 (\pi )$ is defined in Eq.~(\ref{eq:A_1}). 
By substituting these relations into Eq.~(\ref{eq:vector-current-general}), 
the extra vector current can be written in terms of pion fields as
\begin{eqnarray}
V^{a \mu}_3 & = & {1 \over {{\pi} M} }\, {\epsilon}^{abc} \, \hat{\pi}^b 
\, {\partial}_{\nu} {\pi}^m 
\, {\rm sin}^2 \left( { {\pi} \over {f_{\pi}} }
\right) \nonumber \\[5pt]
& & \quad\times 
 \left\{ - {\kappa}_{\pi} \,
{\epsilon}^{cdn} \left( \left[ {\delta}^{dm} - \hat{\pi}^d \hat{\pi}^m
\right] + { {1} \over {A_1( \pi )} } \, \hat{\pi}^d \hat{\pi}^m \right)
\Nbar {\sigma}^{\mu \nu} {\tau}^n N + 4 {\beta}_{\pi} \, {\delta}^{cm}
\Nbar N g^{\mu \nu} \right\} \ . \nonumber \\
& & {}
\label{eq:vector-3-pi}
\end{eqnarray}
For the additional  axial current (\ref{eq:axial-current-general}), 
one obtains 
\begin{eqnarray}
A^{a \mu}_3 & = & {1 \over {f_{\pi} M} } \left\{
{\kappa}_{\pi} \, 
{\epsilon}^{bcd} \, \Nbar {\tau}^d {\sigma}^{\mu \nu} N
- 4 {\beta}_{\pi} \, {\delta}^{bc}\, \Nbar N g^{\mu \nu} \right\} 
{\partial}_{\nu} {\pi}^m \nonumber \\[5pt]
& & \quad \times 
 \left( {\rm cos} \left( { {\pi} \over {f_{\pi}} }  \right) 
\left[ {\delta}^{ab} - 
\hat{\pi}^a \hat{\pi}^b  \right] +
\hat{\pi}^a \hat{\pi}^b \right)
\left\{ A_1( \pi )  \left[ {\delta}^{cm} 
  - \hat{\pi}^c \hat{\pi}^m  \right] +
\hat{\pi}^c \hat{\pi}^m \right\} \ .
\label{eq:axial-3-pi}
\end{eqnarray}

Now one can construct the corresponding total charge
densities used to check the chiral algebra of charges.  
For the vector charge density, one finds
\begin{eqnarray}
V^{a 0} & = & V^{a 0}_{\pi} + V^{a 0}_{\pi N} +V^{a 0}_3 \nonumber \\[5pt]
& = & A_1^2 ( \pi ) \: {\epsilon}^{abc} {\pi}^b 
\left( 1 + { {4 {\beta}_{\pi}} \over {f_{\pi}^2 M} } \, \Nbar N \right)
{\partial}_{0} {\pi}^c +
N^{\dag} { {{\tau}^b} \over 2 } N \:
\left[ {\rm cos} \left( { {\pi} \over {f_{\pi} } } \right) 
\left({\delta}^{ab} - \hat{\pi}^a \hat{\pi}^b \right) 
+ \hat{\pi}^a \hat{\pi}^b \right] \nonumber \\[5pt]
& & \quad {} + 
 g_A \: {\epsilon}^{abc} \: \hat{\pi}^b \: 
{\rm sin} \left({ {\pi} \over {f_{\pi} } } \right) 
N^{\dag} {\gamma}_5 { {{\tau}^c} \over 2} N \nonumber \\[5pt]
& & \quad {} - { {{\kappa}_{\pi}} \over {f_{\pi}^2 M} } \,
{\epsilon}^{abc} \,{\pi}^b A_1^2 (\pi )\,
{\partial}_{i} {\pi}^m {\epsilon}^{cdn}
\left(  \left[ {\delta}^{dm} - \hat{\pi}^d \hat{\pi}^m  \right] +
{ {1} \over {A_1( \pi )} } \, \hat{\pi}^d \hat{\pi}^m \right) 
\Nbar {\sigma}^{0i} {\tau}^n N \ ,
\label{eq:vector-charge-3-pi}
\end{eqnarray}
while the corresponding axial-vector charge density is
\begin{eqnarray}
A^{a 0} & = & A^{a 0}_{\pi} + A^{a 0}_{\pi N} +A^{a 0}_3 \nonumber \\[5pt]
& = & - f_{\pi}^2 
\left( 1 + { {4 {\beta}_{\pi}} \over {f_{\pi}^2 M} }\, \Nbar N \right)
 {\partial}_{0} {\pi}^b
\left\{ {1 \over {\pi}} \, 
{\rm sin} \left( { {\pi} \over {f_{\pi}} } \right)
{\rm cos} \left( { {\pi} \over {f_{\pi} } } \right) 
\left({\delta}^{ab} - \hat{\pi}^a \hat{\pi}^b \right)
+ {1 \over {f_{\pi}} }  \, \hat{\pi}^a \hat{\pi}^b \right\}
\nonumber \\[5pt]
& & \quad {} -  {\epsilon}^{abc} \: \hat{\pi}^b \:
{\rm sin} \left({ {\pi} \over {f_{\pi} } }\right)  N^{\dag} { {{\tau}^c} 
\over 2} N
- {g_A}\mkern2mu N^{\dag} {\gamma}_5 { {{\tau}^b} \over 2 } N 
\left[ {\rm cos} \left( { {\pi} \over {f_{\pi} } } \right) 
\left( {\delta}^{ab} - \hat{\pi}^a \hat{\pi}^b \right) + \hat{\pi}^a
\hat{\pi}^b \right] \nonumber \\[5pt]
&  & \quad {} + { {{\kappa}_{\pi}} \over {f_{\pi} M} } \,
{\epsilon}^{bcd} \, \Nbar {\sigma}^{0i} {\tau}^d N \:
{\partial}_{i} {\pi}^m 
\nonumber\\[5pt]
& & \qquad\quad \times
\left( {\rm cos} \left( { {\pi} \over {f_{\pi}} } \right)  
\left[ {\delta}^{ab} - 
\hat{\pi}^a \hat{\pi}^b  \right] +
\hat{\pi}^a \hat{\pi}^b \right) 
\left\{ A_1( \pi )  \left[ {\delta}^{cm} 
- \hat{\pi}^c \hat{\pi}^m  \right] +
\hat{\pi}^c \hat{\pi}^m \right\} \ .
\label{eq:axial-charge-3-pi}
\end{eqnarray}
The factor $\left( 1 + 4 {\beta}_{\pi} \Nbar N / {f_{\pi}^2 M}  \right)$
in the first term of both densities
represents the net effect of the ${\delta}_2 {\cal L}$ term in the
lagrangian ${\cal L}_{3}$, which is proportional to ${\beta}_{\pi}$.  
It is interesting that the same factor times
${\partial}_{0} {\pi}^a$ can be readily expressed in terms of the pion
canonical momentum, when one inverts the full expression for the
momentum, including the contributions of the new pion--nucleon
terms in the lagrangian.
Thus one finds
\begin{eqnarray}
\Big( 1 & + & 
{ {4 {\beta}_{\pi}} \over {f_{\pi}^2 M} } \, \Nbar N \Big)  
 {\partial}_{0} {\pi}^a \nonumber \\[4pt]
&  & \quad {} = \left[ { 1 \over {A_1^2 (\pi )} }
 \left({\delta}^{ab} - \hat{\pi}^a \hat{\pi}^b \right) + \hat{\pi}^a
\hat{\pi}^b \right] P^b_{\pi}
-  {\epsilon}^{abc}\, {1 \over {A_1^2 (\pi )}}\, 
  {\pi}^b {1 \over {{\pi}^2} } \,
{\rm sin}^2 \left( { {\pi} \over {2 f_{\pi}} } \right) \, 
N^{\dag} {\tau}^c N  \nonumber \\[5pt]
& & \qquad {} - 
 { 1 \over {2 f_{\pi} }} \, g_A \left[ {1 \over{ A_1 (\pi )} }  
\left( {\delta}^{ab}- 
\hat{\pi}^a \hat{\pi}^b \right) + \hat{\pi}^a \hat{\pi}^b \right] 
N^{\dag} {\gamma}_5 {\tau}^b N \nonumber \\[5pt]
& & \qquad {} + 
 { {{\kappa}_{\pi}} \over {f_{\pi}^2} M}
\left( {\epsilon}^{anc} 
  - { {A_1 (\pi ) - 1} \over {A_1 (\pi )} }\, {\epsilon}^{abc} \, 
\hat{\pi}^b \hat{\pi}^n  
+ [A_1 (\pi ) - 1] \, {\epsilon}^{bnc} \, 
  \hat{\pi}^a \hat{\pi}^b \right)
{\partial}_i {\pi}^n \Nbar {\sigma}^{0i} {\tau}^c N \ , 
  \nonumber \\
 & & \null
     \label{eq:full-pi-momentum}
\end{eqnarray}
and the additional lagrangian ${\delta}_2 {\cal L}$ proportional to
${\beta}_{\pi}$ has {\em no effect\/} on the full 
charges, when they are written in terms of canonical momenta.

By substituting the relation (\ref{eq:full-pi-momentum}) into
Eqs.~(\ref{eq:vector-charge-3-pi}) and (\ref{eq:axial-charge-3-pi})
for the vector and axial-vector charge densities, 
and by carrying out the necessary
algebra, one remarkably arrives at precisely the same expressions for
the charge densities in terms of the canonical momenta
[Eqs.~(\ref{eq:charge-2-vector}) and (\ref{eq:charge-2-axial})] as
those obtained with no $\nu = 3$ terms included!  Thus the expressions
for the Noether charges in terms of canonical momenta are not
influenced by the presence of the term proportional to
${\kappa}_{\pi}$ either.  Note, however, that these $\nu = 3$ terms
\textit{will} generally influence the three-vector currents.

Since Eqs.~(\ref{eq:charge-2-vector}) and
(\ref{eq:charge-2-axial}) are known to satisfy the correct
chiral charge algebra, we conclude that the $\nu = 3$ pion--nucleon terms 
in the lagrangian do not influence the charge algebra 
whatsoever.\footnote{%
This conclusion is certainly plausible given the form of the $\nu = 3$
terms in the EFT lagrangian (\protect\ref{eq:eft-lagrangian}).}

\section{interaction amplitudes to order $\ \nu = 3$}
\label{sec:moreamps}

To lowest order in the pion field, the Noether axial current 
(\ref{eq:axial-3-pi}) can be written as
\begin{equation}
A^{a \mu}_{3} \approx { {{\kappa}_{\pi}} \over {f_{\pi} M} }\,
{\epsilon}^{abc} \: {\partial}_{\nu} {\pi}^b \: \Nbar {\tau}^c
{\sigma}^{\mu \nu} N - { {4 {\beta}_{\pi}} \over {f_{\pi} M} } \: \Nbar
N \: {\partial}^{\mu} {\pi}^a + \cdots \ .
\label{eq:axial-3-pi-lowest}
\end{equation}
We calculate additional contributions to the interaction amplitudes
originating from the lagrangians (\ref{eq:delta-1-lagrangian}) and
(\ref{eq:delta-2-lagrangian}) separately.  
We first need to find the corresponding interaction vertices, and we
begin by considering ${\delta}_1 {\cal L}$. 
Recall that to lowest order in the pion field,
\begin{equation}
a_{\nu} \approx { 1 \over {2 f_{\pi}} } \, 
\mbox{\boldmath $ \tau \, \cdot $} \,
{\partial}_{\nu} \mbox{\boldmath $ \pi $} + \cdots  \ ,
\end{equation}
\begin{equation}
v_{\mu \nu} \approx { {1} \over {f_{\pi}^2} } \, {\epsilon}^{abc}\,
{\partial}_{\mu} {\pi}^a {\partial}_{\nu} {\pi}^b \,{ {{\tau}^c} \over
{2} } + \cdots \ .
\label{eq:v-low-order}
\end{equation}
The extra term produces a new lowest-order, strong-interaction vertex
that looks exactly like the one in Fig.~\ref{fig:pi-pi-n0},
but which corresponds now to an analytical expression\footnote{%
Note that Eqs.~(\protect\ref{eq:kappa-vertex}) and
(\protect\ref{eq:beta-vertex}) contain a symmetry factor of 2;
thus, Feynman diagrams including these vertices need to be drawn
only once.}
\begin{equation}
{ {2 i {\kappa}_{\pi}} \over {f_{\pi}^2 M} } \,{\epsilon}^{abc}\, {
  {{\tau}^c} \over {2} } \, k_{\mu} q_{\nu} {\sigma}^{\mu \nu} \ .
   \label{eq:kappa-vertex}
\end{equation}
%

%%%%%%%%%%%%%%%%%%%%%%%%%%%%%%%%%%%%%%%%%%%%%%%%%%%%%%%%
\begin{figure}[htbp]
%\bigskip
\medskip
\begin{center}
\includegraphics*[width=2.2in,angle=0]{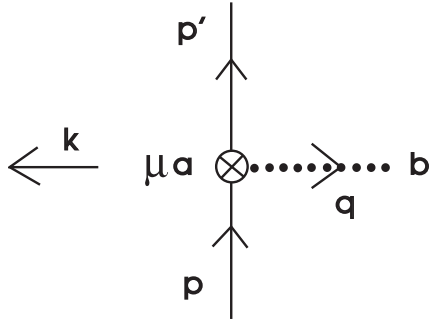}

  \caption{The vertex for the
           one-body axial current arising from 
           ${\delta}_1 {\cal L}$.
           Here $p^{\mu} = q^{\mu} + k^{\mu} + {p'}^{\mu}$.}
      \label{fig:pi-a-n_l1}
\end{center}
\end{figure}
%
%%%%%%%%%%%%%%%%%%%%%%%%%%%%%%%%%%%%%%%%%%%%%%%%%%%%%%%%

The one-nucleon, axial-current interaction vertex due to
${\delta}_1 {\cal L}$ (see Fig.~\ref{fig:pi-a-n_l1}) follows from the
first term in Eq.~(\ref{eq:axial-3-pi-lowest}):
\begin{equation}
{} - { {2{\kappa}_{\pi}} \over {f_{\pi} M} } \, {\epsilon}^{abc} \, q_{\nu}
  {\sigma}^{\mu \nu} \, { {{\tau}^c} \over {2} } \ .
   \label{eq:pi-a-n-vertex} 
\end{equation}
This vertex gives an additional amplitude for axial-current pion production 
on a single nucleon that resembles
the last two diagrams in Fig.~\ref{fig:pi-pi-n} and corresponds to the
expression
\begin{equation}
 M^{ab \: \mu} (\pi )
 = {} - { 1 \over {f_{\pi}} } { {2{\kappa}_{\pi}}
\over { M} } \, {\epsilon}^{abc} \left\{ 
 g^{\mu}_{{\phantom \mu} \lambda} -
{ {k^{\mu} k_{\lambda}} \over {k^2 - m_{\pi}^2} } \right\} \: 
q_\nu \, \bar{u} (p') \, {\sigma}^{\lambda \nu} \, 
{ {{\tau}^c} \over {2} } \, u(p) \ .
   \label{eq:delta1-pi-production}
\end{equation}
This amplitude satisfies PCAC because
\begin{equation}
k_{\mu} \: M^{ab \: \mu} (\pi )
\propto \left\{ k_{\lambda} - { {k^2} \over
{k^2 - m_{\pi}^2} } k_{\lambda} \right\} \, 
%%%%%%%  \Inthelimit
\lower1.9ex\vbox{\hbox{$\  
       \buildrel{\hbox{\Large \rightarrowfill}}
       \over{\scriptstyle{m_{\pi} \, \rightarrow \, 0}}\ $}}
\, 0 \ .
\end{equation}

%%%%%%%%%%%%%%%%%%%%%%%%%%%%%%%%%%%%%%%%%%%%%%%%%%%%%%%%
\begin{figure}[htb]
\medskip
\begin{center}
\includegraphics*[width=4.3in,angle=0]{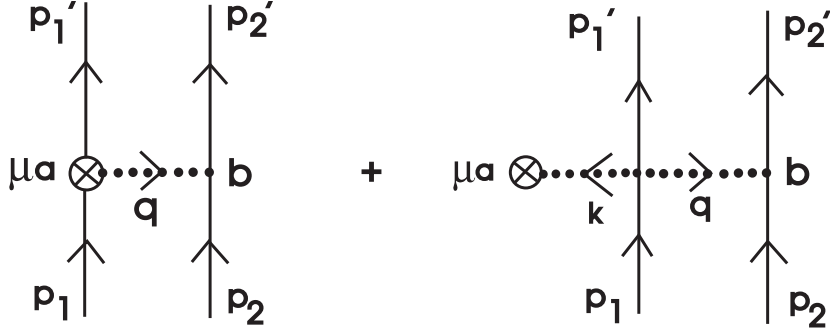}

  \caption{Two-nucleon, axial-current amplitude originating
           from ${\delta}_1 {\cal L}$
           [Eq.~(\protect\ref{eq:delta1-amplitude})].}
      \label{fig:pi-a-2n}
\end{center}
\end{figure}
%
%%%%%%%%%%%%%%%%%%%%%%%%%%%%%%%%%%%%%%%%%%%%%%%%%%%%%%%%
%

By combining all of the previous results, 
one can calculate an additional
two-nucleon, axial-current interaction amplitude due to the
${\delta}_1 {\cal L}$ term (see Fig.~\ref{fig:pi-a-2n}):
\begin{eqnarray}
M^{a \mu} (2)
& = & {}- i \, { {g_A} \over {f_{\pi}^2} } { {2{\kappa}_{\pi}}
\over { M} } \, {\epsilon}^{abc}
\left\{ g^{\mu}_{{\phantom \mu} \lambda} - { {k^{\mu}  k_{\lambda}} 
\over {k^2 - m_{\pi}^2} } \right\} \: \left[
\bar{u} (p_1') {\sigma}^{\lambda \nu} \, { {{\tau}^c} \over {2} } \, u(p_1) 
\right. \nonumber \\
& & \quad {} \times  \left.
 { {q_{\nu} q_{\sigma}} \over {q^2 - m_{\pi}^2} } \:
\bar{u} (p_2') {\gamma}^{\sigma} {\gamma}_5 \, { {{\tau}^b} \over {2} } 
\, u (p_2) + (\mathrm{direct})_2 + \mathrm{cross\ terms} \right] \ .
   \label{eq:delta1-amplitude}
\end{eqnarray}
Here ``$(\mathrm{direct})_2$'' again denotes contributions 
from diagrams where the
axial-current vertex resides on the second nucleon, which can be obtained
from the first term by the replacements 
$p_1 \leftrightarrow p_2$, $p'_1 \leftrightarrow p'_2$, and $q \to Q$,
and ``cross terms'' denotes similar diagrams in which the fermion
lines are crossed in the final state.
This amplitude produces the first additional contribution to the
nuclear AXC originating from the ${\cal L}_{3}$ terms in the EFT
lagrangian.

Now recall that there is yet another term
(\ref{eq:delta-2-lagrangian}) in the $\nu = 3$, pion--nucleon lagrangian
that can be written to lowest order in the pion field as
\begin{equation}
{\delta}_2 {\cal L} \approx {2 \beta_\pi \over f^2_\pi M} \,
\Nbar N \, {\partial}_{\mu} {\pi}^a
{\partial}^{\mu} {\pi}^a \ .
\end{equation}
The corresponding vertex again looks like that in
Fig.~\ref{fig:pi-pi-n0} but now represents an analytical expression
\begin{equation}
- { {4 i {\beta}_{\pi}} \over {f_{\pi}^2 M} }\,
 k_{\mu} q^{\mu} {\delta}^{ab} \ .
   \label{eq:beta-vertex}
\end{equation}
The one-nucleon, axial-current interaction vertex looks the same as in
Fig.~\ref{fig:pi-a-n_l1} and can be deduced from the second term in
Eq.~(\ref{eq:axial-3-pi-lowest}) to be
\begin{equation}
 { {4 {\beta}_{\pi}} \over {f_{\pi} M} } \, 
 q^{\mu} {\delta}^{ab} \ .
    \label{eq:delta2-nucleon}
\end{equation}

It is easy to write out a corresponding axial-current, pion-production
amplitude (again resembling the last two diagrams 
in Fig.~\ref{fig:pi-pi-n}):
\begin{equation}
 M^{ab \: \mu} (\pi )
 = { {4 {\beta}_{\pi}} \over {f_{\pi} M} }
\left\{ q^{\mu} - { {k^{\mu}} \over {k^2 - m_{\pi}^2} } \, k \cdot q
\right\} {\delta}^{ab} \, \bar{u}(p') u(p) \ .
   \label{eq:delta2-pi-production}
\end{equation}
This amplitude also satisfies PCAC:
\begin{equation}
k_{\mu}\, M^{ab \: \mu} (\pi ) \propto \left\{ k \cdot q - { {k^2}
\over {k^2 - m_{\pi}^2} } \, k \cdot q \right\} \, 
%%%%%%%  \Inthelimit
\lower1.9ex\vbox{\hbox{$\  
       \buildrel{\hbox{\Large \rightarrowfill}}
       \over{\scriptstyle{m_{\pi} \, \rightarrow \, 0}}\ $}}
\, 0 \ .
\end{equation}
The corresponding two-nucleon, axial-current interaction 
allows one to identify additional contributions to the axial
two-body amplitude originating from the ${\delta}_2 {\cal L}$ 
term in the lagrangian.
(They also look like Fig.~\ref{fig:pi-a-2n}.)
One obtains
\begin{eqnarray}
 M^{a \mu} (2) = i \, { {g_A} \over {f_{\pi}^2} } { {4 
{\beta}_{\pi}} \over {M} } \left\{ q^{\mu} - { {k^{\mu}} \over {k^2 -
m_{\pi}^2} } \, k \cdot q \right\} { q_{\sigma} \over {q^2 -
m_{\pi}^2} } \: \left[
\bar{u}(p'_1) u(p_1) \: \bar{u} (p_2')
{\gamma}^{\sigma} {\gamma}_5 \, { {{\tau}^a} \over {2} } \, u (p_2) 
\right. \nonumber \\
\hspace{2.7in}
{} + (\mathrm{direct})_2 
 + {} \mathrm{cross\ terms}\, \bigg] \ . \ 
         \label{eq:delta2-amplitude}
\end{eqnarray}
The sum of the amplitudes (\ref{eq:delta1-amplitude}) and
(\ref{eq:delta2-amplitude}) constitutes the total contribution of the
pion--nucleon $\nu = 3$ terms in the EFT lagrangian to the AXC.

\section{rho meson terms in the EFT lagrangian}
\label{sec:rho}

\subsection{Currents and canonical momenta}

An additional piece in the EFT lagrangian (\ref{eq:delta-rho-lagrangian}) 
contains the rho meson field
\begin{equation}
{\rho}_{\mu}(x) \equiv { { 1} \over {2} } \,
\mbox{\boldmath $ \tau \cdot \rho $}_{\mu}(x) \ ,
\end{equation}
which behaves under a chiral transformation as
\begin{equation}
{\rho}_{\mu}'(x) = h(x) {\rho}_{\mu}(x) h^{\dag}(x) \ .
\end{equation}
In principle, one can confirm the same chiral charge algebra as
before to all orders in the pion field.  This follows just as in the
case of the pionic interactions: the contributions of extra pieces in
the lagrangian are absorbed in the modified pion canonical momentum.
To simplify the subsequent equations, however,
we present the proof only to lowest order in the pion field.

For infinitesimal vector transformations,
\begin{equation}
{ {\partial \!\left( \delta {\rho}_{\mu}^c \right)} \over {\partial
{\beta}^a } } = - {\epsilon}^{abc} {\rho}_{\mu}^b \ ,
\end{equation}
while for infinitesimal axial transformations,
\begin{equation}
{ {\partial \!\left( \delta {\rho}_{\mu}^f \right)} \over {\partial
{\alpha}^a } } =  {1 \over {2 f_{\pi}}} \, {\epsilon}^{abc}
{\epsilon}^{cdf} {\pi}^b {\rho}_{\mu}^d \ .
\end{equation}
One also recalls that to lowest order in pion fields,
\begin{equation}
v_{\mu} \approx { {1} \over {2 f_{\pi}^2} } \, {\epsilon}^{abc} {\pi}^a
\, {\partial}_{\mu} {\pi}^b \, { {{\tau}^c} \over 2} \ .
\end{equation}

\begin{figure}[htbp]
\bigskip
\begin{center}
\includegraphics*[width=1.4in,angle=0]{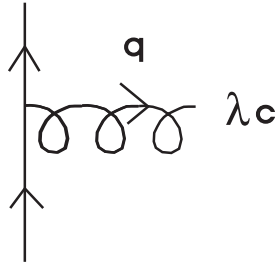}
  \caption{The rho--nucleon vertex from 
           Eq.~(\protect\ref{eq:rho-n}).
           Here $q^{\mu} = p_i^{\mu} - p_f^{\mu}$.}
      \label{fig:rho-n}
\end{center}
\end{figure}

\begin{figure}[htbp]
%\bigskip
\begin{center}
\includegraphics*[width=1.92in,angle=0]{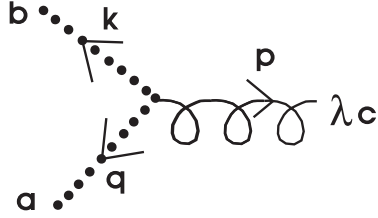}
  \caption{The rho--pi--pi vertex from 
           Eq.~(\protect\ref{eq:rho-pi-pi}).
           Here $p^{\mu} + k^{\mu} + q^{\mu} = 0$.}
      \label{fig:pi-pi-rho1}
\end{center}
\end{figure}

Additional terms in the EFT lagrangian due to the presence of the
rho meson are given by ${\cal L}_\rho$
in Eq.~(\ref{eq:delta-rho-lagrangian}).
To lowest relevant order in the meson fields, one obtains the 
following strong-interaction vertices for the $\rho$ meson:
\begin{equation}
- ig_{\rho} { {{\tau}^c} \over 2 } \left( {\gamma}^{\lambda} + {
{f_{\rho}} \over {2M} } \, i q_{\nu} {\sigma}^{\nu \lambda} \right)
\label{eq:rho-n}
\end{equation}
and 
\begin{equation}
 \,{ {g_{\rho \pi \pi}} \over {m_{\rho}^2} } \, {\epsilon}^{abc} \,
2 ( k \! \cdot \! p \, q^{\lambda} - q \! \cdot \! p \, k^{\lambda})
\ ,
   \label{eq:rho-pi-pi}
\end{equation}
where $k^\lambda$ and $q^\lambda$ are {\em outgoing}
pion momenta, and Eq.~(\ref{eq:rho-pi-pi}) includes a symmetry
factor of 2.
These are shown in Figs.~\ref{fig:rho-n} and
\ref{fig:pi-pi-rho1}.
%
%%%%%%%%%%%%%%%%%%%%%%%%%%%%%%%%%%%%%%%%%%%%%%%%%%%%%%%%
%
%%%%%%%%%%%%%%%%%%%%%%%%%%%%%%%%%%%%%%%%%%%%%%%%%%%%%%%%

Next one can calculate the total additional Noether currents
using the traditional definition [see Ref.~\cite{Ser86},
Eq.~(7.5)]:
\begin{equation}
{\delta} {\cal J}^{\lambda}_a = - { {\partial ({\cal L}_{\rho})} \over
{\partial ({\partial}_{\lambda} \xi)} } { {\partial (\delta \xi)}
\over {\partial {\epsilon}^a} } - { {\partial ({\cal L}_{\rho})} \over
{\partial ({\partial}_{\lambda} {\xi}^{\dag})} } { {\partial (\delta
{\xi}^{\dag})} \over {\partial {\epsilon}^a} } -
{ {\partial ( {\cal L}_{\rho} ) } \over {\partial({\partial}_{\lambda}
 {\rho}^d_{\mu})} } 
{ {\partial ( \delta {\rho}_{\mu}^d )} \over {\partial
{\epsilon}^a } }
\equiv {\delta}_{{\xi}+{\xi}^{\dag}} {\cal J}^{\lambda}_a + {\delta}_{\rho}
{\cal J}^{\lambda}_a \ ,
\end{equation}
where the $\bm{\epsilon}$ are global infinitesimal parameters
($\bm{\alpha}$ or $\bm{\beta}$).
The partial derivative of the lagrangian (\ref{eq:delta-rho-lagrangian})
with respect to the $\rho$ field is
\begin{equation}
{ {\partial ( {\cal L}_{\rho} ) } \over {\partial({\partial}_{\mu}
 {\rho}^d_{\lambda})} } =  - { {f_{\rho} g_{\rho}} \over {4M} } \,
 \Nbar {\sigma}^{\mu \lambda} {\tau}^d N - {\rm Tr} ({\tau}^d
 {\rho}^{\mu \lambda}) \nonumber \\ - g_{\rho \pi \pi} \,{ {2 f_{\pi}^2}
 \over {m_{\rho}^2} } \, {\rm Tr} ({\tau}^d {v}^{\mu \lambda}) \ .
\end{equation}
Thus, to lowest order in the meson fields, the part of the 
Noether vector current
due to differentiating with respect to the $\rho$ field is
\begin{equation}
{\delta}_{\rho} V^{\lambda}_a = - {\epsilon}^{abc} {\rho}^b_{\nu}
 \left\{ { {f_{\rho} g_{\rho}} \over {4M} } \, \Nbar {\sigma}^{\lambda
 \nu} {\tau}^c N + {\partial}^{[ \lambda} \,
{\rho}^{\nu ]}_c + O({\pi}^2 ) \right\} \ ,
\label{eq:vec-rho}
\end{equation}
where the brackets around the superscripts signify the antisymmetric
combination, and repeated isospin indices are summed over, regardless
of their (vertical) position.
The corresponding axial current is
\begin{equation}
{\delta}_{\rho} A^{\lambda}_a = { {1} \over {2 f_{\pi}} }
{\epsilon}^{abc} {\epsilon}^{cdf} {\pi}^b {\rho}^d_{\nu} \left\{
{ {f_{\rho} g_{\rho}} \over {4M} } \, \Nbar {\sigma}^{\lambda \nu}
{\tau}^f N + {\partial}^{[ \lambda} \,
{\rho}^{\nu ]}_f + O({\pi}^2 ) \right\} \ .
\end{equation}

The canonical momentum for the $\rho$ field is, to lowest 
order,\footnote{%
Observe that $(P_\rho )^{0 \, f} = 0$ for this massive vector meson.}
\begin{equation}
(P_{\rho})^{\lambda \, f} \equiv { {\partial ( {\cal L}_\rho ) } \over
 {\partial({\partial}_{0} {\rho}^f_{\lambda})} } = - { {f_{\rho}
 g_{\rho}} \over {4M} } \, \Nbar {\sigma}^{0 \lambda} {\tau}^f N -
 {\partial}^{[ 0}_{\vphantom{f}} \,
{\rho}^{\lambda ]}_f + O({\pi}^2 )\ ,
\end{equation}
which implies
\begin{equation}
{\delta}_{\rho} V^{a 0} =
{\epsilon}^{abc} {\rho}^b_{\nu} (P_{\rho})^{\nu \, c}
\label{eq:vector-rho-momentum}
\end{equation}
and
\begin{equation}
{\delta}_{\rho} A^{a 0} = - { {1} \over {2 f_{\pi}} } \, {\epsilon}^{abc} 
{\epsilon}^{cdf} {\pi}^b {\rho}^d_{\nu} (P_{\rho})^{\nu \, f} \ .
\label{eq:axial-rho-momentum}
\end{equation}

\begin{figure}[bht]
\bigskip
\begin{center}
\includegraphics*[width=2.25in,angle=0]{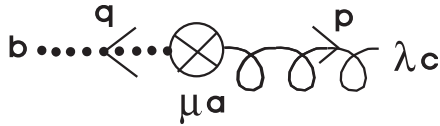}
  \caption{The vertex contribution from the axial current
           (\protect\ref{eq:axial-rho}), as given in
           Eq.~(\protect\ref{eq:rho-pi-axial-vertex}).}

      \label{fig:rho-pi-a}
\end{center}
\end{figure}

Consider now a derivative of the lagrangian
(\ref{eq:delta-rho-lagrangian}) with respect to the $\xi$ field.  
To lowest order in the fields, the corresponding axial current is
\begin{equation}
 {\delta}_{{\xi}+{\xi}^{\dag}} A^{\lambda}_a \approx 
  {} - { {1} \over {2
f_{\pi}} } \, { {f_{\rho} g_{\rho}} \over {4M} } \, {\epsilon}^{abc}
{\epsilon}^{cdf} {\pi}^b {\rho}^d_{\nu} \, 
\Nbar {\sigma}^{\lambda \nu}
{\tau}^f N + g_{\rho \pi \pi} \, { {2 f_{\pi}} \over {m_{\rho}^2} }\,
{\epsilon}^{abc}\, {\partial}_{\nu} {\pi}^b {\partial}^{[ \lambda} \,
{\rho}^{\nu ]}_c \ ,
\end{equation}
and thus the {\em total} additional axial current is
\begin{equation}
{\delta} A^{\lambda}_a =
 g_{\rho \pi\pi} \, { {2 f_{\pi}} \over {m_{\rho}^2} } \, 
{\epsilon}^{abc} \,
{\partial}_{\nu} {\pi}^b \, {\partial}^{[ \lambda} \,
{\rho}^{\nu ]}_c + O(\rho^2 \pi , \, \rho \pi^3 ) \ .
  \label{eq:axial-rho}
\end{equation}
It is now clear why we kept the $\rho\pi\pi$ coupling of order
$\nu = 4$: the $\nu = 3$ contributions to the axial-vector
current cancel out.
The vertex arising from the interaction of this current with the
external source can be calculated as in Eqs.~(\ref{eq:Lext}) and
(\ref{eq:Sext}),
with the result (see Fig.~\ref{fig:rho-pi-a})
\begin{equation}
{} - i 
  { {2 f_{\pi} g_{\rho \pi \pi}} \over {m_{\rho}^2} }\, {\epsilon}^{abc}
\left( {p}^{\mu} {q}^{\lambda} - p \! \cdot \! q \: g^{\mu \lambda} \right)
\ .
  \label{eq:rho-pi-axial-vertex}
\end{equation} 
%
%%%%%%%%%%%%%%%%%%%%%%%%%%%%%%%%%%%%%%%%%%%%%%%%%%%%%%%%
%
%%%%%%%%%%%%%%%%%%%%%%%%%%%%%%%%%%%%%%%%%%%%%%%%%%%%%%%%

Similarly, one finds for the corresponding vector current,
\begin{equation}
 {\delta}_{\xi + {\xi}^{\dag}} V^{\lambda}_a \approx  { 1 \over {2
f_{\pi}^2} } \, { {f_{\rho} g_{\rho}} \over {4M} } \, {\epsilon}^{abc}
{\epsilon}^{cdf} {\epsilon}^{fm\ell} {\pi}^b {\pi}^d {\rho}^m_{\nu} \;
\Nbar {\sigma}^{\lambda \nu} {\tau}^\ell N - { {2 g_{\rho \pi \pi}}
\over {m_{\rho}^2} } \, {\epsilon}^{abc} {\epsilon}^{cdf} {\pi}^b \,
{\partial}_{\nu} {\pi^d} \, \partial^{[\lambda} {\rho}^{\nu]}_f \ . \ 
\label{eq:rho-pi-vector-current}
\end{equation}
These terms are of higher order in the fields, namely
$O(\rho {\pi}^2 )$, than those in Eq.~(\ref{eq:vec-rho}) 
and thus are not considered in the sequel.

\subsection{The corresponding chiral algebra}
\label{sec:algebra}

The terms in the additional $\rho$ meson lagrangian 
(\ref{eq:delta-rho-lagrangian}) that involve derivatives of fields
can be written, to lowest order in the pion field, as
\begin{eqnarray}
\delta {\cal L}_{\partial} & \approx &
 - { {f_{\rho} g_{\rho}} \over {4M} } \, {\partial}_{\mu} {\rho}_{\nu}^a 
\, \Nbar {\sigma}^{\mu \nu} {\tau}^a N
+ { {1} \over {2 f_{\pi}^2} } \,
{ {f_{\rho} g_{\rho}} \over {4M} } \,
{\epsilon}^{abc}  {\epsilon}^{cdf} {\pi}^a \, {\partial}_{\mu}
{\pi}^b {\rho}_{\nu}^d \, \Nbar {\sigma}^{\mu \nu} {\tau}^f N 
         \nonumber \\[4pt]
&  & \quad
  {} - { {1} \over {2} } \,{\partial}_{\mu} {\rho}_{\nu}^a 
\, {\partial}^{ [ \mu} {\rho}^{\nu ] }_a 
- \newg \, {\epsilon}^{abc} {\rho}_{\mu}^a \,
{\rho}_{\nu}^b \, {\partial}^{\mu} {\rho}^{\nu}_c 
- { { g_{\rho \pi \pi}} \over {m_{\rho}^2} } \,
{\epsilon}^{abc} \, {\partial}_{\mu}
{\pi}^a {\partial}_{\nu} {\pi}^b {\partial}^{[\mu} {\rho}^{\nu ]}_c \ .
\end{eqnarray}
Hence
\begin{equation}
{ {\partial (\delta {\cal L}_{\partial}) \over {\partial
( {\partial}_{\lambda} {\pi}^a)} }} = - { {1} \over {2 f_{\pi}^2} }\,
{ {f_{\rho} g_{\rho}} \over {4M} } \, {\epsilon}^{abc} {\epsilon}^{cdf}
{\pi}^b {\rho}_{\nu}^d \, \Nbar {\sigma}^{\lambda
\nu} {\tau}^f N 
- { {2 g_{\rho \pi \pi}} \over {m_{\rho}^2} } \,
{\epsilon}^{abc} \, {\partial}_{\nu} {\pi}^b \,
{\partial}^{[ \lambda} \, {\rho}^{\nu ]}_c \ .
\end{equation}
Moreover, again to lowest order,
\begin{equation}
{\delta}_{\xi + {\xi}^{\dag}} A^{a \lambda} = - f_{\pi} \: {
{\partial (\delta {\cal L}_{\partial}) \over {\partial (
{\partial}_{\lambda} {\pi}^a)} }}
\end{equation}
and
\begin{equation}
{\delta}_{\xi + {\xi}^{\dag}} A^{a 0} = - f_{\pi} 
\: \delta P_{\pi}^a \ .
\end{equation}
This term has the same form as the pionic contribution in
Eq.~(\ref{eq:pi-2-axial-charge}), when the latter is expanded
to leading order in the pion field.
Thus, when the full axial-vector charge density
is expressed in terms of the total pion canonical momentum, 
{\em the pion terms look the same as they did before.}  
This implies that the axial-vector piece 
${\delta}_{\xi + {\xi}^{\dag}} A^{a 0}$ does
not influence the chiral charge algebra. 
One also recalls that the
vector current ${\delta}_{\xi + {\xi}^{\dag}} V^{a \lambda}$
in Eq.~(\ref{eq:rho-pi-vector-current}) is of
higher order in the meson fields and does not influence the 
chiral charge algebra to the order that we consider here.

The only remaining pieces to consider for the proof of the
chiral charge algebra are ${\delta}_{\rho} V^{a0}$ 
and ${\delta}_{\rho} A^{a0}$.  
We can write the total charge densities as
\begin{equation}
V^{a 0}_{\rm tot} \equiv  V^{a 0} + {\delta}_{\rho} V^{a 0} \ ,
\label{eq:total-vector-current}
\end{equation}
\begin{equation}
A^{a 0}_{\rm tot} \equiv  A^{a 0} + {\delta}_{\rho} A^{a 0} \ .
\label{eq:total-axial-current}
\end{equation}
Here $V^{a 0}$ and $A^{a 0}$ are taken from 
Eqs.~(\ref{eq:vector-charge-3-pi}) and (\ref{eq:axial-charge-3-pi})
expanded to lowest order in the meson fields.

First, consider the commutator of the total axial charge densities
\begin{equation}
[ A^{a 0}_{\rm tot}, \: {A^{b 0}_{\rm tot}} ] =[ A^{a 0}, \: A^{b 0}
] + [ A^{a 0}, \: {\delta}_{\rho} A^{b 0} ] + [ {\delta}_{\rho} A^{a
0}, \: A^{b 0} ] + [ {\delta}_{\rho} A^{a 0}, \: {\delta}_{\rho} A^{b
0} ] \ .
\end{equation}
The first commutator in this expression is known:\footnote{%
The two sides of this relation are denoted proportional to each
other because we have omitted the spatial integrations (that
define the charges) and delta functions for brevity.
There are no other numerical factors, and the charges are indeed
normalized correctly.}
\begin{equation}
[ A^{a 0}, \: A^{b 0} ] \propto i {\epsilon}^{abc} V^{c 0} \ ,
\label{eq:ax-ax-inter}
\end{equation}
while the second and third terms are equal to
\begin{equation} 
[ A^{a 0}, \: {\delta}_{\rho} A^{b 0} ] \approx - { {1} \over {2
 f_{\pi}} } \, {\epsilon}^{b\ell c} {\epsilon}^{cdf} (-f_{\pi})
 {\rho}^d_{\nu} (P_{\rho})^{\nu \, f} \left[ P_{\pi}^a , \: {\pi}^\ell
 \right] \ .
\end{equation}
Thus, to lowest order in the meson fields,
\begin{equation}
[ A^{a 0}, \: {\delta}_{\rho} A^{b 0} ] + [ {\delta}_{\rho} A^{a 0},
\: A^{b 0} ] \propto i {\epsilon}^{abc} {\epsilon}^{cdf} {\rho}^d_{\nu}
(P_{\rho})^{\nu \, f} = i {\epsilon}^{abc} {\delta}_{\rho} V^{c 0} \ .
\label{eq:ax-ax-rho}
\end{equation}
The final commutator is of third order in the meson fields,
\begin{equation}
[ {\delta}_{\rho} A^{a 0}, \: {\delta}_{\rho} A^{b 0} ] 
\propto O(\rho {\pi}^2 ) \ ,
\end{equation}
and does not contribute to the charge algebra to the order we
are working.
By combining the results in Eqs.~(\ref{eq:ax-ax-inter}) and
(\ref{eq:ax-ax-rho}), we find that the first relation of the charge
algebra holds in the presence of the additional terms in the lagrangian:
\begin{equation}
[ (Q^a_5 )_{\rm tot}, \: (Q^b_5 )_{\rm tot} ] = i {\epsilon}^{abc}\,
Q^c_{\rm tot} \ .
\end{equation}

Next, consider the commutator of vector charge densities
\begin{equation}
[ V^{a 0}_{\rm tot}, \: {V^{b 0}_{\rm tot}} ] =[ V^{a 0}, \: V^{b 0} ]  
+ [ V^{a 0}, \: {\delta}_{\rho} V^{b 0} ]
+ [ {\delta}_{\rho} V^{a 0}, \: V^{b 0} ]
+ [ {\delta}_{\rho} V^{a 0}, \: {\delta}_{\rho} V^{b 0} ] \ .
\end{equation}
As before, the first commutator is known:
\begin{equation}
[ V^{a 0}, \: V^{b 0} ] \propto i  {\epsilon}^{abc} V^{c 0}  \ ,
\end{equation}
while the second and third commutators vanish because pion,
nucleon, and rho factors commute with each other.
The final commutator can be written as
\begin{equation}
[ {\delta}_{\rho} V^{a 0}, \: {\delta}_{\rho} V^{b 0} ] \approx
{\epsilon}^{adc} {\epsilon}^{bmn}
\left[ {\rho}^d_{\nu} (P_{\rho})^{\nu \, c}, \; {\rho}^m_{\mu}
(P_{\rho})^{\mu \, n} \right] \ ,
\end{equation}
or, after some algebraic manipulation,
\begin{equation}
[ {\delta}_{\rho} V^{a 0}, \: {\delta}_{\rho} V^{b 0} ] \propto i
 {\epsilon}^{abc} {\epsilon}^{cdn} {\rho}^d_{\nu} (P_{\rho})^{\nu \, n}
 = i {\epsilon}^{abc} {\delta}_{\rho} V^{c 0} \ .
\end{equation}
Thus the second relation of the chiral algebra,
\begin{equation}
[ Q^a_{\rm tot}, \: Q^b_{\rm tot} ] 
    = i {\epsilon}^{abc}\, Q^c_{\rm tot} \ ,
\end{equation}
also holds.

Finally, consider the commutator of axial and vector charge densities:
\begin{equation}
[ A^{a 0}_{\rm tot}, \: {V^{b 0}_{\rm tot}} ] =[ A^{a 0}, \: V^{b 0} ]  
+ [ A^{a 0}, \: {\delta}_{\rho} V^{b 0} ]
+ [ {\delta}_{\rho} A^{a 0}, \: V^{b 0} ]
+ [ {\delta}_{\rho} A^{a 0}, \: {\delta}_{\rho} V^{b 0} ] \ .
\end{equation}
The first commutator is known:
\begin{equation}
[ A^{a 0}, \: V^{b 0} ] \propto i {\epsilon}^{abc} A^{c 0} \ .
\end{equation}
The second commutator vanishes because pion and nucleon fields
commute with rho fields.
The third commutator can be written as
\begin{equation}
[ {\delta}_{\rho} A^{a 0}, \: V^{b 0} ] \propto - { i \over {2 f_{\pi}}
}\, {\epsilon}^{a\ell c} {\epsilon}^{cdf} {\epsilon}^{bn\ell} {\pi}^n
{\rho}^d_{\nu} (P_{\rho})^{\nu \, f} \ .
\end{equation}
The final commutator is
\begin{equation}
[ {\delta}_{\rho} A^{a 0}, \: {\delta}_{\rho} V^{b 0} ] \approx - { 1
\over {2 f_{\pi}} } \, {\epsilon}^{af\ell} {\epsilon}^{\ell dc} 
{\epsilon}^{bmn}
{\pi}^f \left[ {\rho}^d_{\nu} (P_{\rho})^{\nu\, c}, \; {\rho}^m_{\mu}
(P_{\rho})^{\mu\, n} \right] \ ,
\end{equation}
which, after some algebra, becomes
\begin{equation}
[ {\delta}_{\rho} A^{a 0}, \: {\delta}_{\rho} V^{b 0} ] \propto - { i
\over {2 f_{\pi}} }\, {\epsilon}^{an\ell} {\epsilon}^{\ell bc} 
{\epsilon}^{cdf}
{\pi}^n {\rho}^d_{\nu} (P_{\rho})^{\nu\, f} \ .
\end{equation}
By using the relation
\begin{equation}
{\epsilon}^{a\ell c} {\epsilon}^{bn\ell} + {\epsilon}^{an\ell} 
{\epsilon}^{\ell bc}
   = {\epsilon}^{abm} {\epsilon}^{mnc}\ , \quad 
\end{equation}
one can show that the sum of the third and fourth commutators is
\begin{equation}
[ {\delta}_{\rho} A^{a 0}, \: V^{b 0} ]
+ [ {\delta}_{\rho} A^{a 0}, \: {\delta}_{\rho} V^{b 0} ] \propto
 i {\epsilon}^{abc} \left( - { {1} \over {2 f_{\pi}} } \, {\epsilon}^{cnp} 
{\epsilon}^{pdf} {\pi}^n {\rho}^d_{\nu} (P_{\rho})^{\nu\, f} \right) \ .
\end{equation}
The term in the parentheses is just ${\delta}_{\rho} A^{c 0}$. 
Thus
\begin{equation}
[ (Q^a_5 )_{\rm tot}, \: Q^b_{\rm tot} ] =  
 i  {\epsilon}^{abc} \, (Q^c_5 )_{\rm tot} \ ,
\end{equation}
and therefore all of the relations constituting the chiral charge algebra 
hold in the presence of the included extra terms in the QHD lagrangian.

\subsection{Two-nucleon, axial-current amplitudes}
%\label{sec:twobody}

\begin{figure}[htb]
\medskip
\begin{center}
\includegraphics*[width=3.75in,angle=0]{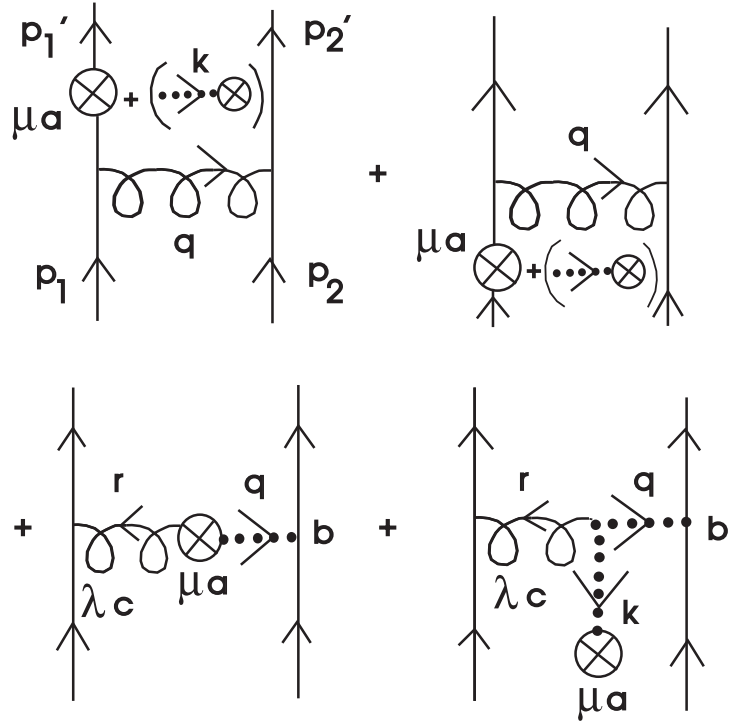}
   \caption{Two-nucleon, axial-current amplitudes containing
            rho meson exchange.
            The notation in the first two diagrams implies that
            the external source can couple to the nucleon line
            in two ways, as in Fig.~\protect\ref{fig:A-1N}.}
       \label{fig:rho-a-2n}
\end{center}
\end{figure}

The two-body, axial-current amplitudes involving the rho meson
are given by the diagrams in Fig.~\ref{fig:rho-a-2n}.
%
%%%%%%%%%%%%%%%%%%%%%%%%%%%%%%%%%%%%%%%%%%%%%%%%%%%%%%%%
%
%%%%%%%%%%%%%%%%%%%%%%%%%%%%%%%%%%%%%%%%%%%%%%%%%%%%%%%%
%
The diagrams in the first row of this figure
originate from the terms of order $\nu \leq 3$ in the $\rho$
meson lagrangian (\ref{eq:delta-rho-lagrangian}), 
while the diagrams in the second row
correspond to the $\nu = 4$ terms.
Thus
\begin{eqnarray}
M^{a \mu}(2) &\approx& i g_A \, g_{\rho}^2 \;
          \frac{(g_{\lambda \alpha} - 
          q_{\lambda}q_{\alpha}/m_{\rho}^2 )}{q^2 - m_{\rho}^2}
 \left(g^{\mu}_{\phantom{\mu}\sigma} - 
          \frac{k^{\mu}k_{\sigma}}{k^2 - m_{\pi}^2} \right) 
                                 \nonumber \\[5pt]
& &  \qquad {}
 \times  \bar{u}(p_2^{\prime})\left(\gamma^{\alpha} - \frac{f_{\rho}}{2M}
   \, i q_{\beta}\sigma^{\beta \alpha} \right) \frac{\tau^c}{2} \, u(p_2) 
                                 \nonumber \\[6pt]
& &  \qquad {}
\times \bar{u}(p_1^{\prime}) \left\{ 
\gamma_5 \gamma^{\sigma} \frac{1}{(\not \! p_1 
                - \stroke{q})-M}
    \left( \gamma^{\lambda} 
    + \frac{f_{\rho}}{2 M} \, i q_{\nu} \sigma^{\nu \lambda} \right)
     \frac{\tau^a}{2} \frac{\tau^c}{2} \right.
                                   \nonumber \\[6pt]
& & \qquad\qquad\qquad\quad  \left.
   {} + 
  \left( \gamma^{\lambda} 
     + \frac{f_{\rho}}{2 M} \, i q_{\nu}        
  \sigma^{\nu \lambda} \right) \frac{1}{(\not \! p^{\mkern2mu \prime}_1 
     + \stroke{q})-M} \,
       \gamma_5 \gamma^{\sigma} \,
     \frac{\tau^c}{2} \frac{\tau^a}{2} \right\} u(p_1)
                                     \nonumber \\[6pt]
& & \quad {}
+ \frac{2 g_A \, g_{\rho} \, g_{\rho \pi \pi}}{m_{\rho}^2} \, 
    \epsilon^{abc} \, \bar{u}(p_2^{\prime})
    \gamma_5 \, \stroke{q} \frac{\tau^b}{2} \,
          u(p_2) \; \bar{u}(p_1^{\prime})
   \left( \gamma^{\lambda}-\frac{f_{\rho}}{2M} \,
     i r_{\nu} \sigma^{\nu \lambda} \right)
            \frac{\tau^c}{2} \, u(p_1)
                                     \nonumber \\[6pt]
& & \qquad {} \times
 \frac{(g_{\lambda \alpha} 
    - r_{\lambda}r_{\alpha}/m_{\rho}^2)}{r^2-m_{\rho}^2}
 \, \frac{1}{q^2-m_{\pi}^2} 
                                    \nonumber \\[5pt]
& & \qquad {} \times
  \left\{ r^{\mu} q^{\alpha} 
   - (r \cdot q)g^{\mu \alpha} +\frac{k^{\mu}}{k^2 - m_{\pi}^2}
  \, [(q \cdot r) k^{\alpha} - (k \cdot r) q^{\alpha} ] \right\}
                                      \nonumber \\[6pt]
& & \quad {} + (\mathrm{direct})_2 + {\mathrm{\ cross\ terms.}}  
   \label{eq:a-2n-rho-amplitude}
\end{eqnarray}
The terms proportional to $g_\rho^2$ 
satisfy PCAC by themselves, precisely as in the nonlinear
realization of the Sigma model. 
The terms proportional to $g_{\rho \pi \pi}$ also
satisfy PCAC separately because
\begin{equation}
 k_{\mu}\left\{ r^{\mu} q^{\alpha} - 
       (r \cdot q)g^{\mu \alpha} +\frac{k^{\mu}}{k^2 - m_{\pi}^2}
  \, [(q \cdot r)k^{\alpha} - (k \cdot r) q^{\alpha} ] \right\} 
\lower1.9ex\vbox{\hbox{$\  
       \buildrel{\hbox{\Large \rightarrowfill}}
       \over{\scriptstyle{m_{\pi} \, \rightarrow \, 0}}\ $}}
\, 0 \ .
\end{equation}
One can identify the corresponding additional AXC from these
amplitudes.

\section{Contributions from sigma and omega mesons}
\label{sec:sigma-omega}

Now one can easily add the $\sigma$ and $\omega$
contributions in the present EFT.
The corresponding lowest-order interaction terms can be deduced from
the full lagrangian (\ref{eq:eft-lagrangian}):
\begin{equation}
\label{eq:sigma-lagrangian}
{\cal L}^{\rm int}_{\sigma} \approx g_s \phi \, \Nbar N 
+ {\eta}_{\rho} { {g_s} m^2_{\rho} \over M}\,  \phi \,
{\rm Tr} ( {\rho}_{\mu} {\rho}^{\mu} ) 
- { {\kappa_3} \over {3!}} \, { {g_s} m^2_{s} \over M}\, {\phi}^3
\ .
\end{equation}
The final two terms involve interactions
that are of higher order in the meson fields than we consider here
for the two-body axial current, so we will not discuss them further.
Similarly,
\begin{equation}
{\cal L}^{\rm int}_{\omega} \approx - g_v V_{\mu} \,
\Nbar {\gamma}^{\mu} N 
- { {f_v g_v} \over {4M} } \, {\partial}_{\,[ \mu} V_{\nu ]}
\mkern2mu \Nbar {\sigma}^{\mu \nu} N
+ {1 \over 2} \, {\eta}_{1} { {g_s} m^2_{v} \over M}\, \phi 
V_{\mu} V^{\mu} \ .
\label{eq:omega-lagrangian}
\end{equation}
Again, the final term above is of higher order
in the meson fields and does not contribute here.

We obtain the following new strong-interaction vertices
that look exactly like the one in Fig.~\ref{fig:pi-n},
but with the pion line replaced by sigma and
omega lines, respectively 
(see Ref.~\cite{Ser86}, Fig.~29):
\begin{equation}
i g_s \ ,
     \label{eq:sigma-n}
\end{equation}

\begin{equation}
-i g_v \left( {\gamma}^{\lambda} + 
{ {f_v} \over {2M} } \, i q_{\nu} {\sigma}^{\nu \lambda} \right)
\ . \\
     \label{eq:omega-n}
\end{equation}
Both $\sigma$ and $\omega$ mesons are chiral scalars and thus do
not contribute to the Noether axial-vector current.

The amplitudes for the two-body, axial-vector currents involving
$\sigma$ and $\omega$ meson exchange are represented by diagrams 
analogous to those in the first row of Fig.~\ref{fig:rho-a-2n}.
They are given by the first term in
Eq.~(\ref{eq:a-2n-rho-amplitude}), with the appropriate
meson propagator substituted for the $\rho$ propagator, and
with the expression (\ref{eq:rho-n}) for the rho--nucleon vertex
replaced by the sigma--nucleon vertex (\ref{eq:sigma-n})
or the omega--nucleon vertex (\ref{eq:omega-n}), respectively.

\section{Summary}
\label{sec:summary}

In this work, we compute the axial-vector current
based on a recently proposed hadronic lagrangian with a
nonlinear realization of chiral symmetry \cite{Fur97}.
The effective lagrangian provides a systematic framework for
calculating both nuclear exchange currents and nuclear
wave functions.
The lagrangian is truncated by working to a fixed order in
the parameter $\nu$, which essentially counts powers
of ratios of particle momenta to the nucleon mass $M$ or
of mean meson fields to the nucleon mass~\cite{Fur97,Rus97}.
Practically speaking, in the nuclear many-body problem,
the expansion is in powers of
$k_{\mathrm{F}}/M$, where $k_{\mathrm{F}}$ is the
Fermi wavenumber at equilibrium nuclear density;
this ratio provides a small parameter for ordinary nuclei and
for electroweak processes at modest momentum transfers.

The present framework has several advantages:
First, because of the nonlinear realization of the chiral symmetry, 
the axial coupling constant $g_A \!\approx\! 1.26$ 
appears naturally as a parameter in the lagrangian.
Second, the explicit enforcement of the
symmetry ensures that the axial current is conserved
in the chiral limit (and that PCAC holds for finite pion mass),
and that the familiar $SU(2)_L \times SU(2)_R$ 
chiral algebra is satisfied by the vector and axial-vector charges.
The symmetry also implies that there will be 
axial-vector exchange currents (AXC) involving nonlinear
meson couplings.
Third, since the same degrees of freedom are used to describe the
axial-vector current and the nuclear dynamics, 
the parameters of the
theory can be calibrated using empirical nuclear and hadronic
properties (or two-nucleon bound-state and scattering data),
together with pion--nucleon scattering observables.
Thus there are no unknown constants in the axial-current
amplitudes.
To our knowledge, these desirable properties have not been
included simultaneously in earlier models of the AXC.

The axial currents are derived here by keeping all relevant terms
in the lagrangian through order $\nu = 3$ (and $\nu = 4$ in
the $\rho$ meson case).\footnote{%
The $\Delta$ resonance can also be included as an explicit degree 
of freedom in the EFT lagrangian, in a manner that maintains
chiral symmetry~\protect\cite{Tang96,Ell98}.
An explicit $\Delta$
would modify our expressions for the currents and covariant
amplitudes, but it would not change the mean-field results for
even-even nuclei~\protect\cite{Fur97,Ser97}.
We leave the explicit inclusion of the $\Delta$ for a future
project.}
In the chiral limit,
the correct chiral charge algebra is proved explicitly to all orders
in the pion field for terms involving pions and nucleons, and to
lowest order in the pion field for terms involving pions, nucleons,
and rho mesons.
For finite pion mass, it is also shown that PCAC holds for the 
one- and two-body axial-current amplitudes, as well as for the 
amplitude for pion production on a single nucleon.
The AXC can be deduced from the two-nucleon amplitudes, although we
do not derive them in this paper. 
[We do, however, compute the leading (nonrelativistic) correction
to the axial charge density in Eq.~(\ref{eq:nu2-axial-charge}).]

Since our analysis of the axial-vector current in the nuclear
many-body problem is performed by splitting our EFT 
lagrangian into numerous pieces, it is useful to summarize here
our most important results and expressions.
The complete vector and axial-vector currents are given in 
Eqs.~(\ref{eq:vector-current}) and (\ref{eq:axial-current}),
plus (\ref{eq:vector-currentpiN}) and
(\ref{eq:axial-currentpiN}) [or
(\ref{eq:V-pi-N}) and (\ref{eq:A-pi-N})],
plus (\ref{eq:vector-current-general}) and
(\ref{eq:axial-current-general}) [or
(\ref{eq:vector-3-pi}) and (\ref{eq:axial-3-pi})],
plus (\ref{eq:vec-rho}) and (\ref{eq:axial-rho}).
Thus, for example, the complete currents are
\beqa
V_{\mathrm{tot}}^{a \mu} & = &
  - i\,{ {f_{\pi}^2} \over 4} \, {\rm Tr} 
\left\{ {\tau}^a \left( U {\partial}^{\mu} U^{\dag} + U^{\dag}
{\partial}^{\mu} U \right) \right\} \nonumber \\[3pt]
 & & \quad
 {} + { {1} \over {4} }\, \Nbar {\gamma}^{\mu} 
\left[ {\xi} {\tau}^a  {\xi}^{\dag} + {\xi}^{\dag} {\tau}^a 
 {\xi} \right] N
+ { {1} \over {4} }\, g_A \Nbar {\gamma}^{\mu} {\gamma}_5
\left[ {\xi} {\tau}^a  {\xi}^{\dag} - {\xi}^{\dag} {\tau}^a 
 {\xi} \right] N \nonumber\\[3pt]
 & & \quad
 {} - { {i {\kappa}_{\pi}} \over {2 M} } \: \Nbar
{\sigma}^{\mu \nu} \left[ {\xi}^{\dag} {\tau}^a \xi - \xi {\tau}^a
{\xi}^{\dag}, \: a_{\nu} \right] N - { {2 {\beta}_{\pi}} \over {M} }
\: \Nbar N \: {\rm Tr} \left[ ( {\xi}^{\dag} {\tau}^a \xi - \xi
{\tau}^a {\xi}^{\dag} ) \: a^{\mu} \right] \nonumber \\[5pt]
 & & \quad
 {} - {\epsilon}^{abc} {\rho}^b_{\nu}
 \left\{ { {f_{\rho} g_{\rho}} \over {4M} } \, \Nbar {\sigma}^{\mu
 \nu} {\tau}^c N + {\partial}^{[ \mu} \,
{\rho}^{\nu ]}_c + O({\pi}^2 ) \right\} 
\eeqa
%@@@
and
\beqa
A_{\mathrm{tot}}^{a \mu} & = &
  - i\,{ {f_{\pi}^2} \over 4}\, {\rm Tr} 
\left\{ {\tau}^a \left( U {\partial}^{\mu} U^{\dag} - U^{\dag}
{\partial}^{\mu} U \right) \right\} \nonumber \\[3pt] 
 & & \quad 
 {} - { {1} \over {4} } \,\Nbar {\gamma}^{\mu} 
\left[ {\xi} {\tau}^a  {\xi}^{\dag} - {\xi}^{\dag} {\tau}^a 
 {\xi} \right] N
- { {1} \over {4} }\, g_A \Nbar {\gamma}^{\mu} {\gamma}_5
\left[ {\xi} {\tau}^a {\xi}^{\dag} + {\xi}^{\dag} {\tau}^a 
 {\xi} \right] N \nonumber \\[3pt]
 & & \quad
 {} - { {i {\kappa}_{\pi}} \over {2 M} } \: \Nbar {\sigma}^{\mu \nu}
\left[ {\xi}^{\dag} {\tau}^a \xi + \xi {\tau}^a {\xi}^{\dag}, 
\: a_{\nu} \right] N
- { {2 {\beta}_{\pi}} \over {M} } \: \Nbar N \: {\rm Tr}
\left[ ( {\xi}^{\dag} {\tau}^a \xi 
+ \xi {\tau}^a {\xi}^{\dag} ) \: a^{\mu} \right] \nonumber \\[5pt]
 & & \quad
 {} + \left\{
   g_{\rho \pi\pi} \, { {2 f_{\pi}} \over {m_{\rho}^2} } \, 
{\epsilon}^{abc} \,
{\partial}_{\nu} {\pi}^b \, {\partial}^{[ \mu} \,
{\rho}^{\nu ]}_c + O(\rho^2 \pi , \, \rho \pi^3 ) \right\} \ .
\eeqa
We also derived expressions for the following amplitudes:
the scattering of a nucleon by an external source
[Eqs.~(\ref{eq:axial-current-nucleon}),
(\ref{eq:pi-a-n-vertex}), and
(\ref{eq:delta2-nucleon})],
one-pion production by an external source
[Eqs.~(\ref{eq:piprod-2}), (\ref{eq:delta1-pi-production}),
and (\ref{eq:delta2-pi-production})],
and nucleon--nucleon scattering in the presence of an external 
source [Eqs.~(\ref{eq:L2-axial-2N-amplitude}),
(\ref{eq:delta1-amplitude}), (\ref{eq:delta2-amplitude}),
and (\ref{eq:a-2n-rho-amplitude})].
All of these results were evaluated at the ``tree'' level,
with no pion loops.
This is because we computed the scattering matrices to lowest
order in the external source $S^{\mathrm{ext}}_{a \mu}$, and we
expanded the interaction lagrangian to leading orders in the
pion field.
As noted above, since our expansion proceeds in powers of pion
momenta (or $m_{\pi}$) relative to the nucleon mass $M$ (or
a similar ``heavy'' mass scale), these tree-level expressions
are valid for modest external pion momenta and momentum
transfers between nucleons.
Our results become exact in the soft-pion limit
(with $m_{\pi} = 0$).

At order $\nu = 2$, our EFT lagrangian generates familiar 
pion-exchange contributions to the AXC.  
These are the same as those calculated by Kubodera, Delorme, 
and Rho \cite{Kub78} in the current-algebra approach.  
We reiterate the cause of the difficulty in 
Ref.~\cite{Ana98}, and how that problem is resolved by the EFT 
lagrangian~(\ref{eq:eft-lagrangian}): in a 
{\em linear\/} realization of the $SU(2)_L \times SU(2)_R$ 
chiral symmetry, the nucleon axial-vector coupling $g_A$ is 
constrained to be {\em unity}.  
Changing this coupling ``by hand'' is equivalent to rescaling 
the fields in the theory and leads to a chiral charge algebra 
that is incorrect.  
In contrast, in a {\em nonlinear\/} realization of the symmetry, 
$g_A$ becomes a free parameter, which allows the 
Goldberger--Treiman relation, PCAC, and the correct chiral 
charge algebra to be satisfied simultaneously.  
This is explicit evidence that chiral symmetry 
is realized {\em nonlinearly\/} in low-energy QCD.

In future work, we will show that the dominant contributions to the
axial-vector current come from one- and two-body amplitudes 
involving pion exchange and are thus model independent~\cite{Rho91}.
The relevant set of $\nu = 3$ AXC will be written in covariant form
for use with relativistic, mean-field Dirac wave functions, and
the dominant $\nu = 3$ terms will also be given in nonrelativistic
form for use with more traditional (e.g., harmonic oscillator) nucleon
wave functions.
We now have the effective lagrangian and corresponding Noether
currents, and the problem is therefore well defined;
however, as emphasized in Refs.~\cite{Dmi98,Ana98}, any
{\em application\/} to the many-body problem must proceed
{\em consistently\/} in terms of wave functions, the interactions
that determine those wave functions, and the current operators
to be used within that framework.

We also plan to derive the meson-exchange corrections to the 
electromagnetic current that are implied by this effective 
lagrangian~\cite{Fur97}.
These meson-exchange corrections can then be used to compute
selected electroweak processes in nuclei.
This will allow, for example, for an investigation of the following
interesting issue:
It is known that calibrating the relevant parameters to nuclear properties
using mean-field nuclear wave functions corresponds to a density-functional
approach \cite{Ser97,Fur97,Ser02}.
In this approach, 
bulk and single-particle nuclear observables are used to define a
set of quasiparticle, single-nucleon wave functions, which implies that
exchange and correlation corrections are (approximately) 
included implicitly in the parameters.
Within this quasiparticle, single-nucleon framework, we expect that the
PCAC relations derived here will remain valid.
In the calculation of exchange-current amplitudes, however, one samples
{\em two-nucleon} wave functions inside the nucleus.
It remains to be seen whether the calibration procedure described above
leads to realistic results for two-body, exchange-current matrix elements,
or if some more complicated calibration procedure (that also includes 
two-body observables) must be used.

\section*{Acknowledgments}

We thank our colleagues R. J.~Furnstahl and J.~Piekarewicz 
for useful comments.
This work was supported in part by the Department of Energy under
Contract Nos.~DE-FG02-87ER40365 and DE-FG02-97ER41023.
\finalnewpage  % if necessary to avoid widows or orphans

\end{document}